\documentclass[twocolumn]{aastex63}
\usepackage{amsmath}
\usepackage{amssymb}
\usepackage{graphicx}
\usepackage{color}
\usepackage{float}

\definecolor{darkgreen}{rgb}{0,0.35,0}
\accepted{by ApJ}

\shorttitle{Transition to Runaway Gas Accretion After Pebble Isolation}
\shortauthors{Chen et al.}

\begin{document}

\title{The Preservation of Super Earths and the Emergence of Gas Giants after
Their Progenitor Cores have Entered the Pebble Isolation Phase}

\author{Yi-Xian Chen}
\affiliation{Department of Physics, Tsinghua University, Beijing, China}

\author{Ya-Ping Li}
\affiliation{Theoretical Division, Los Alamos National Laboratory, Los Alamos, NM, USA}

\author{Hui Li}
\affiliation{Theoretical Division, Los Alamos National Laboratory, Los Alamos, NM, USA}

\author{Douglas N. C. Lin}
\affiliation{Department of Astronomy, University of California, Santa Cruz, CA, USA}
\affiliation{Institute for Advanced Studies, Tsinghua University, Beijing, China}
\correspondingauthor{Yi-Xian Chen}
\email{yx-chen17@mails.tsinghua.edu.cn}



\begin{abstract}
The omnipresence of super-Earths suggests that they are able to be retained in natal disks around low-mass stars, whereas exoplanet's mass distribution indicates that some cores have transformed into gas giants through runaway gas accretion at $\gtrsim 1$ AU from solar-type stars. In this paper, we show that transition to runaway gas accretion by cores may be self-impeded by an increase of the grain opacity in their envelope after they have acquired sufficient mass (typically $\sim 10 M_\oplus$) to enter a pebble-isolation phase. The accumulation of $\sim$mm-m size pebbles in their migration barriers enhances their local fragmentation rates. The freshly produced sub-mm grains pass through the barrier, elevate the effective dust opacity and reduce the radiative flux in the core's envelope. These effects alone are adequate to suppress the transition to runaway accretion and preserve super-Earths in the stellar proximity ($\sim$0.1 AU), albeit entropy advection between the envelope and the disk can further reduce the accretion rate. At intermediate distance ($\sim 1$AU) from their host stars, the escalation in the dust opacity dominates over entropy advection in stalling the transition to runaway accretion for marginally pebble-isolated cores. Beyond a few AU, the transformation of more massive cores to gas giants is reachable before severe depletion of disk gas. This requirement can be satisfied either in extended disks with large scale height via orderly accretion of migrating pebbles or through the mergers of oligarchic protoplanetary embryos, and can account for the correlated occurrence of long-period gas giants and close-in super-Earths.
\end{abstract}



\section{Introduction} \label{sec:intro}
Super Earths are planets with typical radii of 1-4 $R_{\oplus}$, masses of $\sim$10$M_{\oplus}$ at distances of 0.05-0.3AU or 
beyond. They are commonly discovered by radial velocity and transit surveys in 20-30\% of solar systems 
\citep{2010Sci...330..653H,2013ApJ...766...81F,2014ApJ...784...45R}. 
Another rich population of exoplanets with similar masses (A.K.A. sub Neptunes) have been
found by microlensing surveys at larger distance \citep{2012ARA&A..50..411G,2012RAA....12..947M}. 
Although super Earths or sub Neptunes are more difficult to find at distances beyond the snow line, 
there is no shortage of known intermediate and long-period (years) {giants and sub-Jovians}, especially around
metal-rich solar type stars \citep{2010Sci...330..653H,2011Natur.473..349S,2012A&A...544A...9V,2019AJ....158..187B,Suzukietal2018}.  
The presence of embedded super Earths have also been inferred from the structure of protostellar
disks (PPDs) imaged by ALMA \citep{2013ApJ...778...53D,2016ApJ...820L..40A}.
While numerous mechanisms have been proposed to the origin of each population, a unified scenario is still 
needed to concomitantly account for dynamical and structural properties of both super Earth 
and gas giants. Some outstanding issues include 1) the relative occurence rates of super Earths and
gas giants \citep{2010Sci...330..653H}; 2) the limited mass of super Earths' initial atmosphere \citep{OwenWu2017}; 3) similar mass, radii, and separation 
between multiple super Earths \citep{Weissetal2018, Wu2019}; and 4) the correlated association between long-period gas giants and close-in 
super Earths around common host stars \citep{2018AJ....156...92Z,Bryanetal2019}.

\subsection{Super Earth Formation via Core Accretion}
The widely-adopted, orderly-accretion scenario for planet formation is based on the assumption that giant planets form 
after the emergence of super-Earth cores in gas-rich protostellar disks \citep{1996Icar..124...62P, 2004ApJ...604..388I}. 
The germination of cores proceeds either through the merger of oligarchic protoplanetary embryos and planetesimals 
or independently through the monolithic acquisition of inwardly-migrating, optimum-size (mm-m) grains, commonly dubbed as 
pebbles. 

If planetesimals are the main building blocks, they would rapidly
coagulate until a few oligarchic embryos emerge with \textit{embryo-isolation} masses (a fraction to a few $M_\oplus$)
which consume all the planetesimals in their feeding zone \citep{1998Icar..131..171K, 2004ApJ...604..388I}.  Through 
their interaction with each other and with their natal disks, these oligarchs can cross each other's orbit, collide, 
and merge during or after the depletion of gas in the disk \citep{2013ApJ...775...42I}. 

{In cases where coagulation of planetesimals alone is not sufficient to generate a core massive enough to significantly 
perturb the surface density and pressure distribution of the disk gas,} the core's growth can be further 
supplied by fast migrating optimum-size grains (i.e. $\sim$mm-m-size pebbles) \citep{2010A&A...520A..43O}, until it 
reaches the pebble-isolation mass typically a few to $10M_{\oplus}$.  Subsequently, it induces local maxima in the 
surface density and pressure distribution for the disk gas.  This perturbed structure blocks the pebbles' inward 
flow and suppresses the cores' growth
\citep[for a review]{2012A&A...544A..32L,2014A&A...572A..35L,2018A&A...612A..30B,2017ASSL..445..197O}. 

The emergence of pebble-isolated cores is followed by a quasi steady accretion of gas at a rate which is 
determined by the near thermal equilibrium between energy released from accretion, radiation transfer through 
the envelope, and heat exchange with the natal disk \citep{1996Icar..124...62P}.  The rate of gas accretion
increases with the total mass of the core and envelope.  When the Gaseous 
envelope to solid Core mass Ratio (GCR) reaches $\sim 1$, {the heat loss rate is sufficiently high to 
offset the thermal equilibrium in the gaseous envelope and runaway gas accretion is initiated.} If this 
critical GCR can be reached on a ``runaway timescale" before the severe depletion of the disk gas, the cores 
would transform into gas giants. Otherwise, the failed cores would be preserved as super Earths 
\citep{1974Icar...22..416P, 1980PThPh..64..544M, 1980E&PSL..50..202M, 1982P&SS...30..755S, 2004ApJ...604..388I}.

{Numerical calculation of cooling models \citep{2000ApJ...537.1013I,2006ApJ...648..666R,2014ApJ...786...21P,2014ApJ...797...95L,2015ApJ...800...82P} show 
that a typical super Earth candidate of $\sim$10$M_{\oplus}$ within a passive disk environment undergoes transition to runaway 
accretion on a typical timescale of a few Myr, less than or equal to typical disk lifetime \citep{2014prpl.conf..475A}.} Such high transition probability is inconsistent with the ubiquity of super Earths with
masses $\sim 10 M_\oplus$. Indeed \citet{2019MNRAS.488L..12N} concluded that gas accretion rate on to planets needs to be suppressed by about an order of magnitude to match the observed planet mass distribution from ALMA.
 One remedy is that higher opacity would significantly reduce the cooling efficiency, slow down the contraction 
of the envelope, increase the critical mass for transition to runaway gas accretion, and prolong the accretion timescale. For example, 
semianalytic calculation of core accretion model by \citet{2000ApJ...537.1013I} showed that $\tau \propto \kappa$ where 
$\tau$ is the runaway timescale and $\kappa$ is an opacity constant throughout the atmosphere. In these models, grains provide the dominant opacity sources in the outer radiative region where the gas temperature is below the grains' sublimation threshold. Runaway 
accretion might be avoided in ``dusty" disks with metallicity an order of magnitude larger than the solar value \citep{2014ApJ...797...95L,2015ApJ...811...41L}. It was also proposed that the final mass-doubling of cores may have occurred relatively late within disk lifetime \citep{2016ApJ...817...90L}. Such a scenario may encounter challenges from the apparent occurrence correlation between close-in super Earths and long-period gas giants around common host stars \citep{2018AJ....156...92Z}. \citet{2017ApJ...850..198Y} found tidally-forced turbulence to be a potential mechanism that prolongs the accretion timescale, while another study by \citet{2019ApJ...887L..33K} suggested that the pressure at the core surface maybe high enough to sequester the atmosphere into magma core, quenching further growth of the envelope. 

In most early models, spherically symmetric unidirectional infall with a radiative boundary condition is assumed at 
the interface between the envelope and the disk.  In two dimensional planar geometry, gas continually flows between the cores' 
Hill sphere and the background disk. Mixing of gas between these regions also leads to entropy advection such 
that the flow in the outermost region of the envelope is forced to be bidirectional and adiabatic. Although
this thermal structure is insensitive to the opacity, entropy advection alone reduces the efficiency of 
heat transfer and hinders the transition to runaway accretion \citep{2015MNRAS.446.1026O,2015MNRAS.447.3512O}.  
This effect is also conspicuous in some recent 3D simulations of Circumplanetary Disks considering rotation which confirm that 
it unlikely for runaway accretion to be triggered within the disk lifetime \citep{2019ApJ...887..152F}, contrary to previous 1D results. According to a 1D approximate analysis of entropy advection process \citep[hereafter ACL20]{2020MNRAS.494.2440A}, its effectiveness depends on the thermal stratification, especially the entropy at the midplane 
of the disk.  The inner ($\lesssim 1$ AU) regions of protostellar disk are primarily heated by viscous dissipation
whereas the outer region is heated by stellar irradiation \citep{2007ApJ...654..606G}.  In the proximity of the star ($\sim 0.1$ AU), the time delay due to advection of entropy carried by the viscously heated gas is more than two orders 
of magnitude longer than that derived based on passive (irradiated) minimum mass nebula.  
The slow down factor and its dependence on the disk thermal stratification decreases with the distance from the host star. At outer regions $r \gtrsim 1$ AU, the cores' Hill radii becomes much larger than their physical 
size, the time delay in the onset of runaway gas accretion due to entropy advection is limited to a factor of a 
few and it is less sensitive to the disk structure. This result is consistent with an earlier 3D simulation by \citet{2013ApJ...778...77D} at large radii.

\subsection{Environment Impact due to  Pebble Isolation}

The previous core-growth simulations used values from idealistic unperturbed disks at a range of orbital radii for the planet's boundary
conditions.  Therefore opacity of the disk and the envelope is scaled with a solar composition. We explore the 
possibility that this assumption may be modified after pebble isolation is induced by sufficiently massive cores, 
whether this mass is acquired mainly from planetesimal coagulation or pebble accretion.
Their tidal perturbation leads to partial gaps with surface density/pressure maxima in the nearby disk regions
\citep{1980ApJ...241..425G,1986ApJ...309..846L} which influence the motion of solids in the disks and create dust barriers \citep{PaardekooperMellema2006,2006MNRAS.373.1619R}. A gap-opening planet could trap nearly all dust particles $\gtrsim 0.1$mm outside the planet's orbit and generate rings and cavities similar to those identified in transition disks and
ALMA images \citep{2012ApJ...755....6Z,2014ApJ...789...59O,2017ApJ...843..127D,2018ApJ...868...48K}.  Modest planet mass
($M_{\rm Iso} \sim 10 M_\oplus$) is needed to trap inwardly migrating optimum-size (a few mm to decameters) 
grains (commonly dubbed as \textit{pebbles}) outside the orbit \citep{MorbidelliNesvorny2012,2014A&A...572A..35L,2018ApJ...854..153W}. When the progenitor core reaches the pebble isolation mass $M_{\rm Iso}$, it blocks the larger-size ($\sim$mm-m-size) 
pebbles completely in the dust barrier, while the smaller-size (sub-mm) particles that are well-coupled with gas slip through the barrier.  A fraction of the small particles is accreted onto the envelope around the cores. 

The onset of pebble isolation effectively quenches cores' growth via further pebble accretion, setting off the steady accretion of gas at a rate determined by both opacity and entropy advection (see above discussion). For typical
temperature at the outer boundary of the envelope, the dominant sources of opacity are $\mu$m-size grains.
Although these small grains are well coupled to the gas, the ratio of their surface density to the gas surface
density is not the same as that in the unperturbed disk.  The opacity in the envelope and immediate disk vicinity 
may diverge from previously assumed values because a) for the dust distribution that directly influence accretion at 
location of the cores, spiral density waves leads to considerably higher concentration of grains than the 
\textit{azimuthally-averaged} value, and b) the previous simulations didn't incorporate full dust coagulation and 
fragmentation among the pebbles trapped and accumulated in the barrier, as a means to freshly produce sub-mm-size 
grains which can flow to the vicinity of the cores.

In passive outer disks heated by irradiation, this change in the dust density may significantly affect the cores' 
gas accretion rate through opacity enhancement. In contrast, the planet's dynamic perturbation only slightly changes the gas density at the accretion boundary. In active inner disk regions heated by viscous dissipation and cooled by convection 
or radiation \citep{2007ApJ...654..606G}, the elevated opacity also modifies the thermal stratification and midplane 
entropy, adding to the complexity of this process.

In this work we will address these issues and analyze the effect of pebble isolation on gas accretion of progenitor cores, mainly through influence of opacity. In \S\ref{sec:2}, we use the LA-COMPASS codes to run simulations of passively irradiated PPDs for implications on how pebble isolation might change the dust size distribution around the planet from unperturbed disk values.  In \S\ref{sec:3} we apply the hydrodynamic results in opacity expressions of core-growth models to a) determine their effect on accretion timescale of a growing super Earth in passively irradiated disk regions, and b) estimate how similar effects influence super Earths located in actively accreting regions with high entropy.  Both of these analysis are carried out for close-in orbital radii of 0.1-1AU. After showing that the opacity effect quenches gas accretion very effectively even at outer regions (at 5 and 10 AU), we offer a discussion on how gas giant progenitors may reach runaway transition beyond 1 AU in \S\ref{sec:4}. In \S\ref{sec:5} we summarize our results.

\section{From Pebble Isolation to Dust Size Distribution}
\label{sec:2}

\subsection{Hydrodynamical Model}
In this section, we explore how the pebble isolation process changes the dust size distribution around the vicinity of 
the planet by performing two-fluid (gas+dust) hydrodynamic simulations of 2D Protoplanetary Disks (PPDs) with LA-COMPASS code \citep{2005ApJ...624.1003L,2009ApJ...690L..52L}. In general, a larger mass is required for planets to create a pressure bump in a 3D disk. However, \citet{2018A&A...612A..30B} shows that this enhancement is only by a small factor of $\sim 1.5$. Although the Hill radius of a $10M_{\oplus}$ rocky core is smaller than the disk scale height, since most mm-m size pebbles settle into the midplane, the mass associated with a pebble isolation core in a 2D PPD is similar to the 3D case. For a set of fiducial illustrative models, we choose a passive irradiative model in which the PPD's temperature is independent of the distance above the midplane and is $T_{disk} \propto r^{-1/2}$  and its aspect ratio

\begin{equation}
h=\dfrac{c_s}{v_K} (r)=\dfrac {H(r)} r=h_0 \left( {\dfrac{r}{r_p}} \right)^{1/4}
\end{equation}
where $H$ is the scale height of the gas and $r_p=1AU$ is the planet's circular orbital radius. The central star has a mass of 1 $M_{\odot}$, and the planet's mass is set to $10M_{\oplus}$. The disk has an initial gas profile of 

\begin{equation}
\Sigma_g(r)= \Sigma_0 \left( {\dfrac {r}{r_p}} \right)^{-1.5},
\label{eq:fiducial}
\end{equation}
where $\Sigma_0
$ is the surface density of gas at the orbital radius. We adopt a thin disk where $\Sigma_g(1 \mathrm{AU})= 75 \mathrm{g}
\cdot \mathrm{cm}^{-2}$, which is $\gtrsim 20$ times smaller than the commonly-used minumum mass solar nebula (MMSN) and
minimum mass extrasolar nebula (MMEN) models 
\citep{2004ApJ...604..388I,2013MNRAS.431.3444C} due to consideration of computational expenses, as in usual hydrodynamic simulations of PPDs
\citep[e.g.][]{2017ApJ...843..127D,2018ApJ...868...48K}. However the dynamic 
problems in planet-disk interactions is generic as long as disk self-gravity is not too important and the dust density is
normalized by the gas density, therefore the results in low surface density simulations can also be applied to realistic 
environments and provide implications for planet evolution such as pebble isolation 
\citep{2018ApJ...868...48K,2018A&A...612A..30B}, ring formation \citep{2017ApJ...843..127D}, gap opening 
\citep{2015MNRAS.448..994K} and planet migration \citep{2018ApJ...861..140K}.

The dust is set with an initial dust-to-gas ratio of 0.01 corresponding to ISM \citep[e.g.][]{2013AJ....146...19L}, and 
the turbulent viscosity parameter \citep{1973A&A....24..337S} of the gas is $\alpha_g=10^{-3}$. The simulation domain is 
from 0.4AU to 5AU, and carried out with a grid resolution of $n_r\times n_{\phi}=1024 \times 1024$. This will give us a 
thin disk with negligible self-gravity compared to its host star.  We choose $h_0=0.03$ which corresponds to 
$T_{disk} =200$K at 1 AU. The midplane density scales as $\rho_g=\Sigma_g/2H\propto r^{-3}$.

At the inner and outer boundaries, the gas and dust velocities are set to be those in a steady state.
The surface densities of the gas and dust are also set so that the mass flux is constant. A damping method 
is used to avoid artificial wave reflection \citep{2006MNRAS.370..529D,2018ApJ...868...48K} and keep the boundary 
symmetric.

\subsection{Coagulation Model}

The coagulation scheme of LA-COMPASS is introduced in \citet{2019ApJ...878...39L,2019ApJ...885...91D}, incorporating the interaction equations from \citet{2014ApJ...795L..39F} and coagulation model from \citet{2010A&A...513A..79B}. We use 151 species of dust with characteristic size $s_i$ logarithmically spaced between 1 $\mu $m up to 1m sized pebble, the dust distribution is initialized with MRN distribution \citep{1977ApJ...217..425M}, with the dust turbulent viscosity parameter $\alpha_{d}=10^{-2}$. {The main reason for using a different $\alpha$ for dust is that the viscosity of dust is determined by the mid-plane turbulence associated with local instability, while $\alpha_g$ controls the global viscosity for the disk accretion, which may be determined by other MHD processes \citep{Bai_Stone2013}. A larger $\alpha_{\rm d}$ can avoid an extremely large dust size due to coagulation. More detailed reasons are described in \citet{2017ApJ...839...16C,2020ApJ...889L...8L, Lietal2020}.} LA-COMPASS treats each species of dust as a kind of fluid, and to determine the evolution of the dust size distribution we explicitly integrate the Smoluchowski equation in each spatial cell. Turbulence and radial drift are considered as sources of relative velocity distribution between the dust particles \citep{2007A&A...466..413O}. Collisions with relative velocity above threshold velocity $v_{\rm f}=10$m/s result in fragmentation, and those below result in coagulation. {The fragmentation velocity $v_{\rm f}$ depends on the composition of the dust grains. Typical fragmentation velocities for silicate grains measured both theoretically and experimentally are of the order of a few ${\rm m\ s^{-1}}$ (e.g., \citealt{Guttler2010}; also see review by \citealt{Blum2008}). Icy grains have a larger fragmentation velocity and can grow to larger sizes than the silicate grains. The fragmentation velocity could be $v_{\rm f}=10\ {\rm m\ s^{-1}}$ if there is more than 1\% of ice in the mixture \citep[e.g.,][]{Gundlach2015}. We have tested that an even larger $v_{\rm f}$ (i.e., $30\ {\rm m\ s^{-1}}$) results in most of dust particles being more efficiently trapped in the ringed structures, and therefore the increase of the dust surface density in the vicinity of the planet becomes less significant.} The fragmentation process produces fragments that obey a power law mass spectrum $n(m) \mathrm{d} m \propto m^{-1.83} \mathrm{d} m $ \citep{2008A&A...480..859B,2010A&A...513A..79B}. Due to high computing expenses, we calculate the coagulation outcomes every 50 timesteps.

\subsection{Dust Production at the Pebble Barrier}

\begin{figure}[htp!]
\centering
\includegraphics[width=0.48\textwidth]{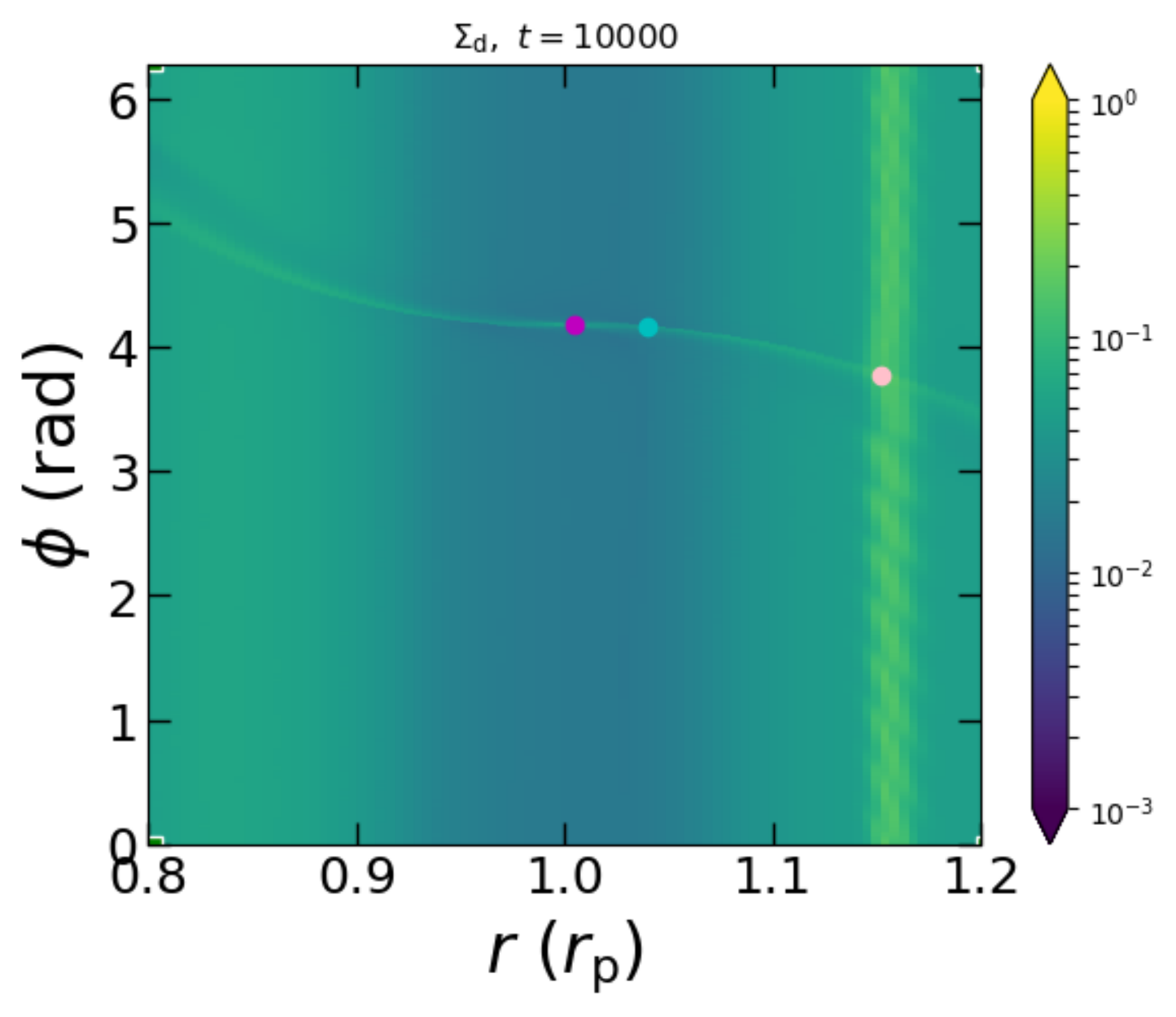}
\caption{Surface density of dust $\Sigma_d$ obtained in the coagulation run after 10000 orbits. The three colored dots represents locations where we trace the evolution of dust size distribution, corresponding to colored lines in Fig \ref{fig:DSD}.}
\label{fig:locations}
\end{figure}

We run the simulations with and without the planet for 10000 orbital periods and neglect disk self gravity and the planet's total mass growth (which is much slower compared to our simulation orbital times) throughout the process. We trace the evolution of dust size distribution at three locations along the density wave as shown in Fig \ref{fig:locations}, which shows the profile for total dust density at $t=10000$. Magenta point represents the direct location of the planet, and the pink point represents a location inside the ring generated by pebble isolation.

\begin{figure*}[htp!]
\centering
\includegraphics[width=0.42\textwidth]{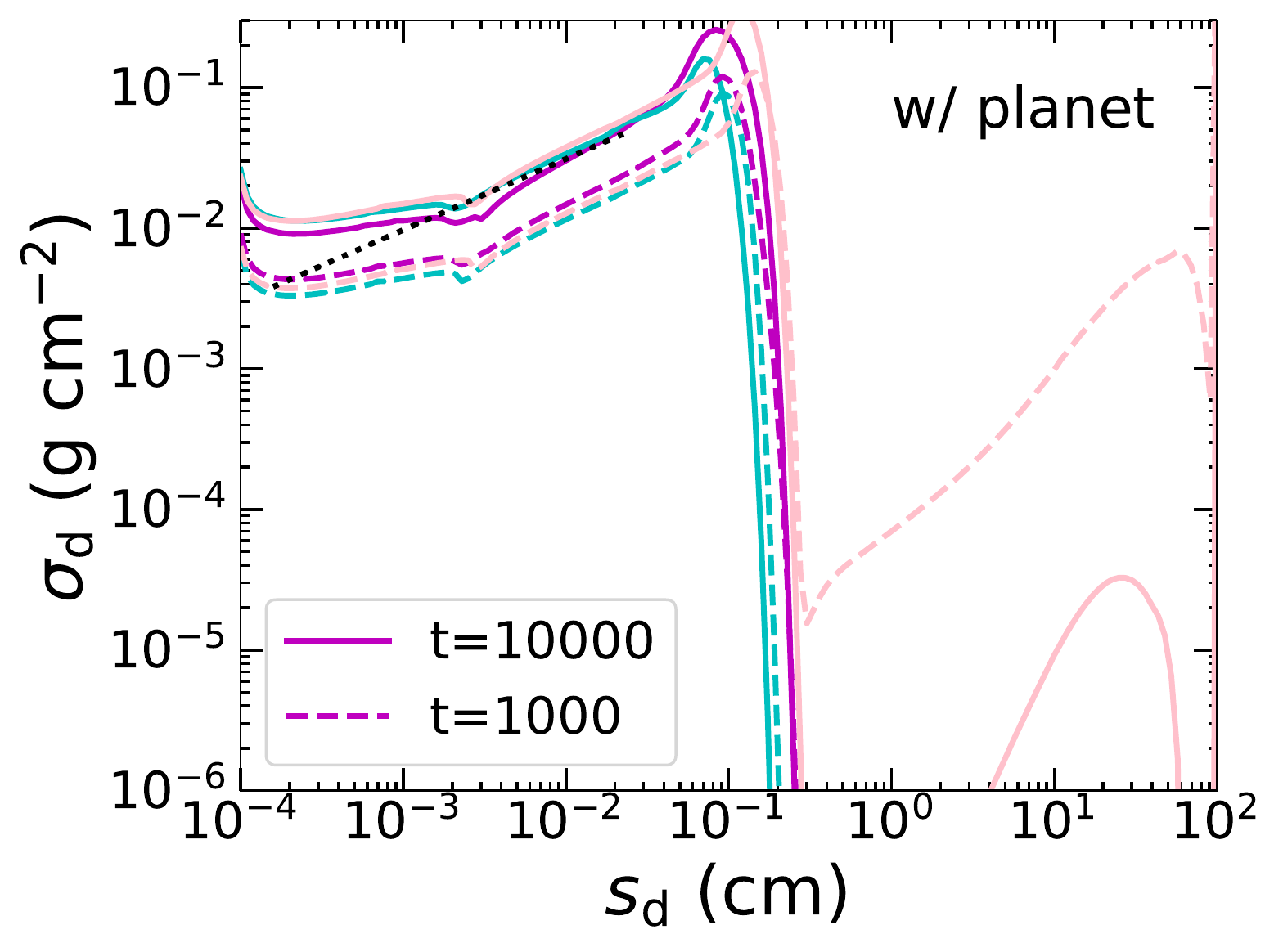}
\includegraphics[width=0.42\textwidth]{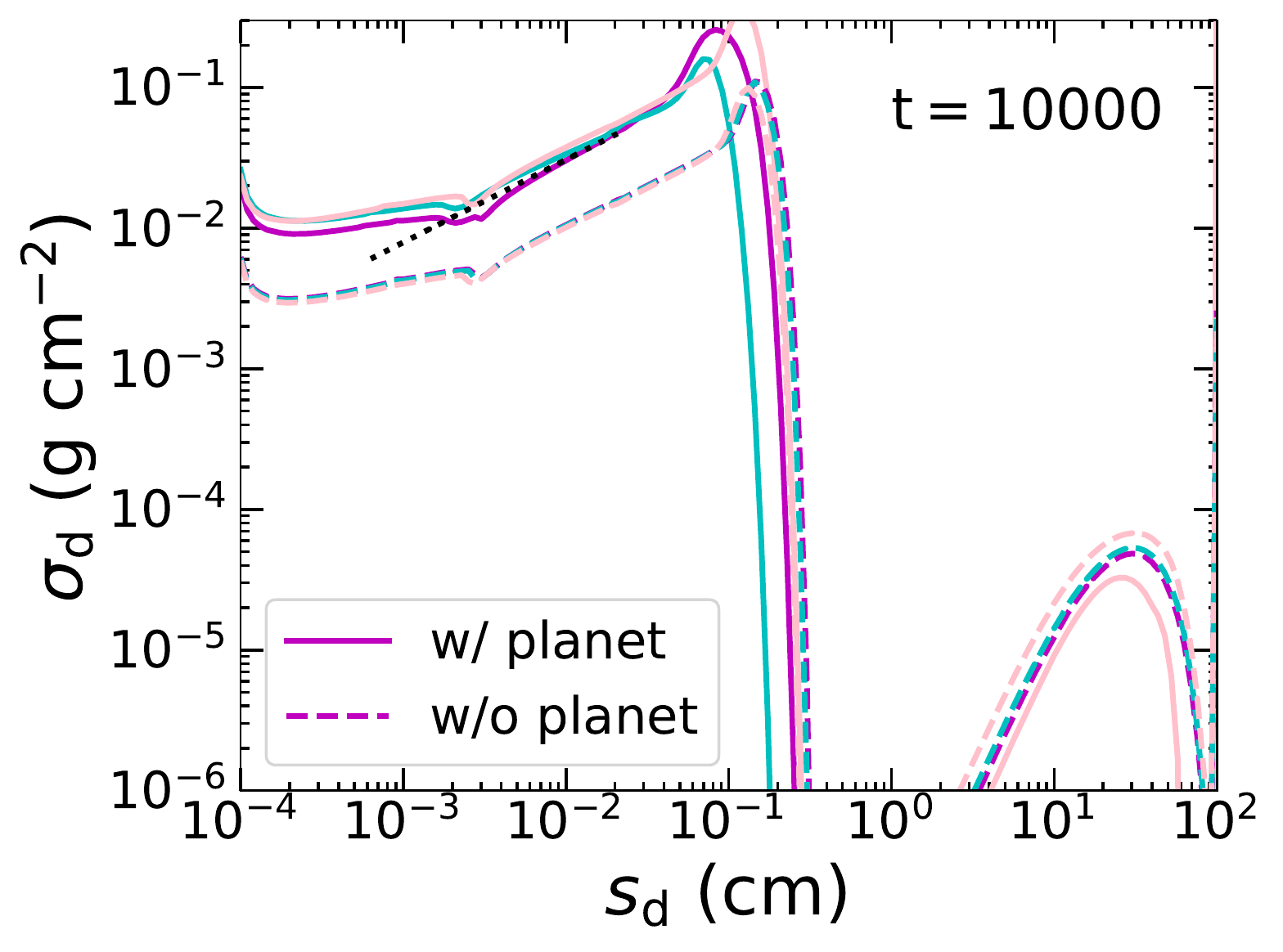}
\caption{Left panel: the dust size distribution (dust surface density per bin) at three locations (Fig \ref{fig:locations}) in the simulation with planet perturbation, dashed lines for 1000 orbits and solid lines for 10000 orbits; Right panel: the dust size distribution at 10000 orbital times for simulation with (solid lines) and without (dashed lines) the planet, the former is constantly growing in magnitude while the latter has already reached a steady state. The black dashed lines correspond to the power-law fit with a slope of $\sim 0.5$ close to MRN distribution.}
\label{fig:DSD}
\end{figure*}

Fig \ref{fig:DSD} shows the comparison between the dust size distributions at the three locations for the unperturbed and perturbed case at 10000 orbits, and between those at $t=1000$ and $t=10000$ in the perturbed case. For the unperturbed and symmetric case (i.e. in the absence of an embedded planet), the growth of dust density stops after the overvall dust-to-gas ratio evolves to $\left.\Sigma_d/\Sigma_g\right|_{final}\approx 0.02$. For the perturbed case (with an embedded planet), the ring at the pressure maxima blocks out pebbles larger than $\sim$0.2 cm leaving the planet location devoid of pebbles, and the dust-to-gas ratio evolves to 0.05. However, the evolution for the perturbed case does not stall at 10000 orbits despite that the shape of dust size distribution has already reached a coagulation-fragmentation equilibrium. 

\begin{figure*}[htp!]
\centering
\includegraphics[width=0.7\textwidth]{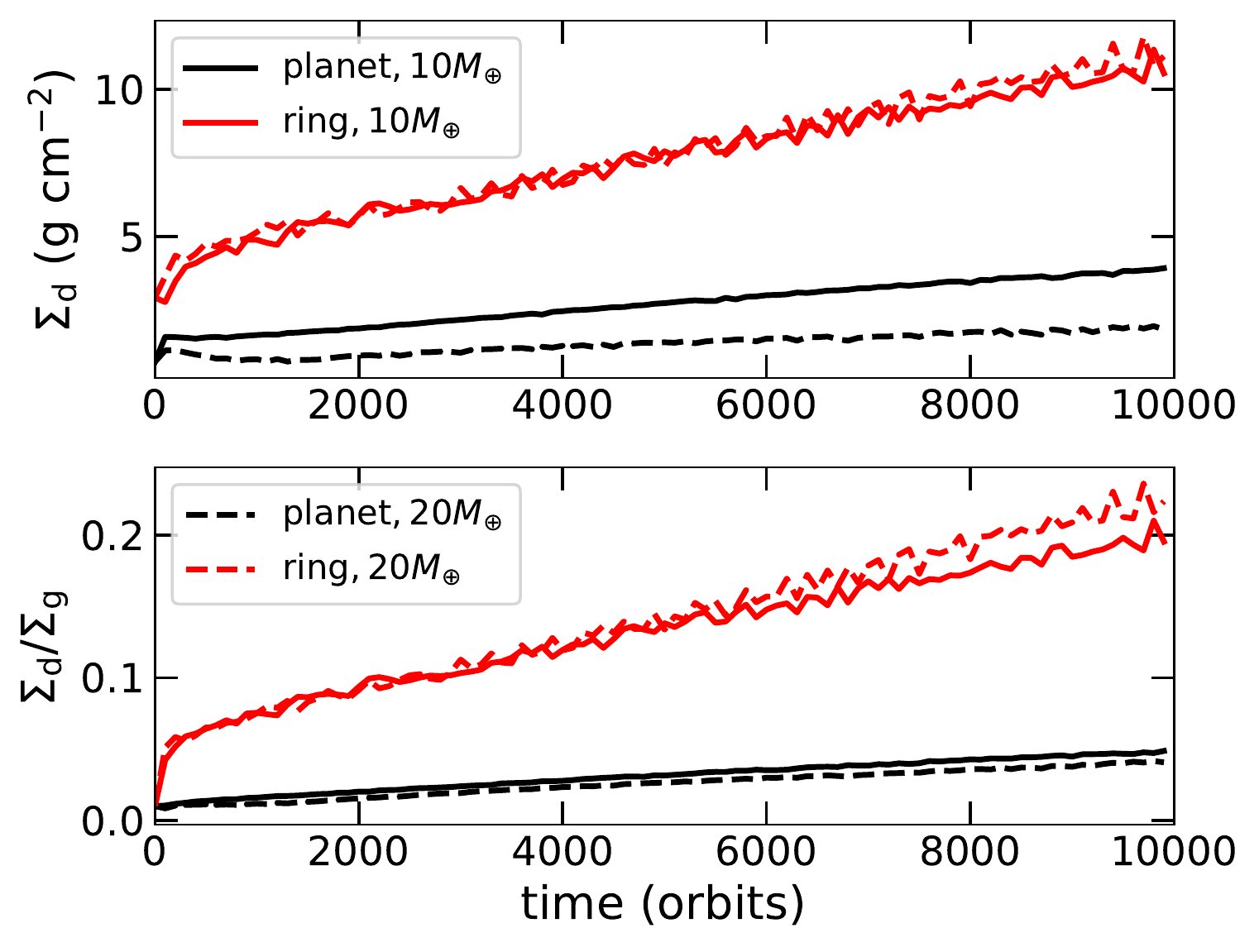}
\caption{The evolution of dust density and dust-to-gas ratio at location of the planet (magenta point in Fig \ref{fig:locations}) and the pebble barrier (pink point \ref{fig:locations}), with respect to orbital time. Bold lines correspond to the $10M_{\oplus}$ case, while dashed lines correspond to $20M_{\oplus}$ case which we discuss in \S \ref{sec:4}.}
\label{fig:lineargrowth}
\end{figure*}

The bold lines in Fig \ref{fig:lineargrowth} shows the dust densities around the $10 M_{\oplus}$ planet (magenta point) and at the location of the pebble barrier (pink point) are increasing linearly with time. The collision and fragmentation in the growing pebble barrier is still replenishing the inner regions with dust. When we extrapolate the profiles, we find that the dust-to-gas ratio inside the pebble barrier will reach $\sim1$ in $\approx$ 70000 orbits which is small compared to the disk lifetime. If we assume that this is going to be the limit for dust-to-gas ratio since afterwards the dust feedback and other instabilities will stall the growth of dust density at the pressure maxima \citep{2004ApJ...608.1050G,2005ApJ...620..459Y,2014ApJ...780...53C,2018ApJ...868...48K,2020arXiv200106986H}, then the overall dust-to-gas ratio at the location of the planet will reach a final value of $\left.\Sigma_d/\Sigma_g\right|_{final} \approx 0.28$ for the planet-perturbed disk, which is $\approx 14$ times larger than the steady value of the unperturbed case, and made up of smaller particles.

\begin{figure*}[htp!]
\centering
\includegraphics[width=0.72\textwidth]{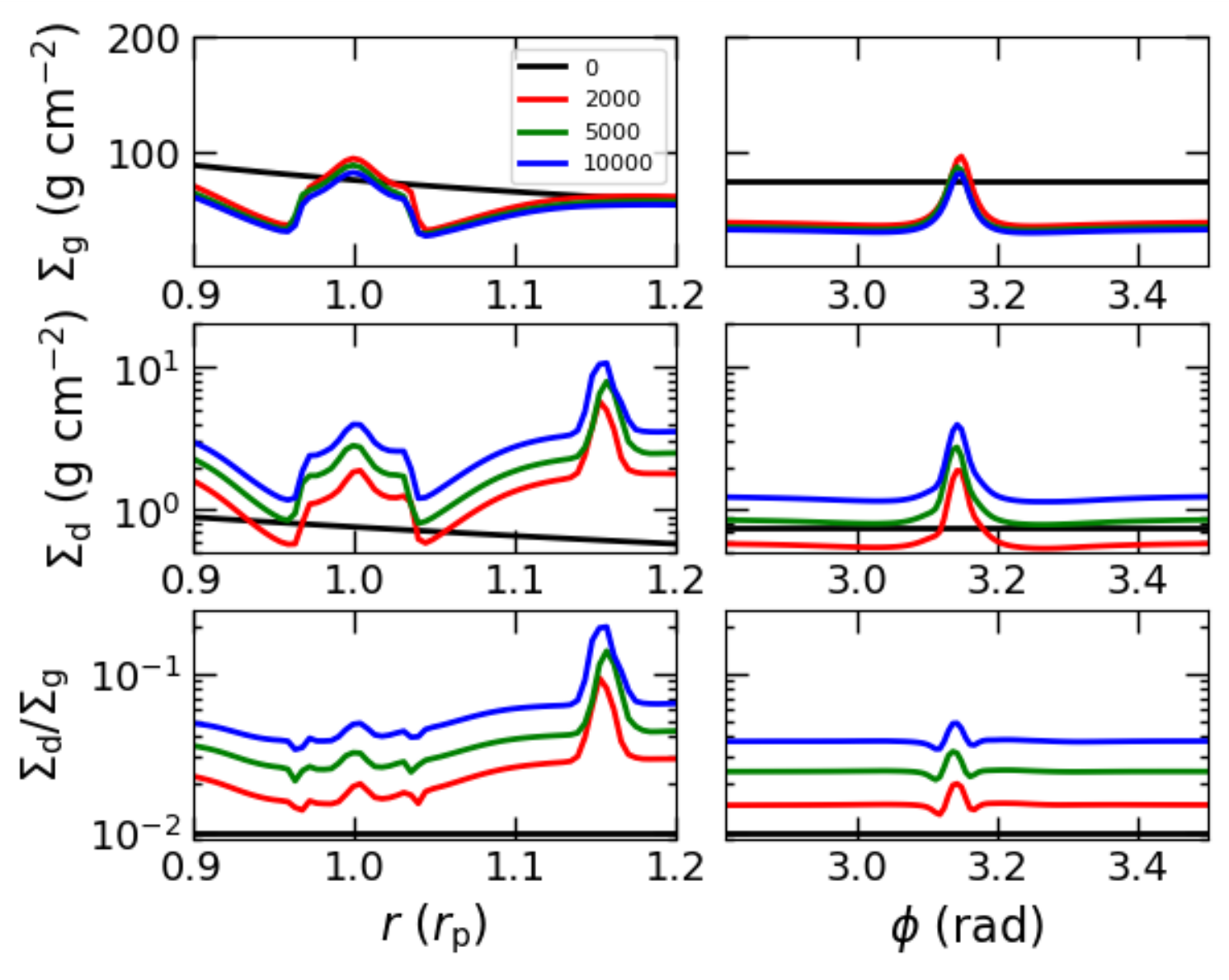}
\caption{Left panel: \textit{radial} distribution of dust, gas density and dust-to-gas ratio at the planet's 
\textit{azimuth}; Right panel:  \textit{azimuthal} distribution of dust, gas density and dust-to-gas ratio at the orbital
radius of the planet $r_p=1$AU. Profiles taken at different orbital times are shown in different colors.}

\label{fig:atplanet}
\end{figure*}

In Fig \ref{fig:atplanet} we plot out some dust, gas density and dust-to-gas ratio \textit{azimuthal profiles} at the 
\textit{radial} location of the planet, and \textit{radial profiles} at the \textit{azimuthal} location of the planet, 
taken at different orbital. For this set of model parameters, partial clearing of the gas distribution leads to shallow 
depression in the $\Sigma_g$ radial distribution near the planet (at $0.96 r_p$ and $1.04 r_p$) and a local maximum 
outside (at $1.2 r_p$) the orbit of the planet.  At planet location $\Sigma_g$ does not differ significantly from the 
unperturbed values, since the gas density reaches a maxima in the azimuthal distribution around the direct vicinity 
planet and cancels out the azimuthally-averaged gap formation effect. Therefore we could assume the boundary condition of gas density for planet 
accretion is similar for perturbed disk and unperturbed disk. In fact, the accretion is very insensitive to this parameter \citep[e.g.][]{2014ApJ...797...95L} that even a change of $\sim$100 in gas density itself is not going to affect core accretion very much. Here we emphasize the size distribution of dust adundance as the dominating factor that influences the runaway timescale. 

\subsection{Different Location and Gas Surface Density}
We have also done simulations for a 10$M_{\oplus}$ core at 0.5AU, and another at 1AU with doubled surface density.
In those higher surface density runs, the concentration of sub-mm grains near the planets are uniformally enhanced at the same rate relative to unperturbed disks, as the fiducial model. This result is 
consistent with pebble isolation phenomenon being generic. 

The only change is the truncation dust size ($\sim s_\mathrm{max}$) in the $\sigma_d-s_d$ distribution at the location of the planet (Fig \ref{fig:DSD}, right panel planet case). Since the Stokes number scales as \citep{2002ApJ...581.1344T}

\begin{equation}
    S t=\frac{\pi}{2} \frac{s_d \rho_{\bullet}}{\Sigma_{\mathrm{g}}},
\label{stokes}
\end{equation}
where $\rho_{\bullet}$ is the internal density of the grains and $s_d$ is the size, we expect the dust species at the truncation radius to have Stokes number given by the fragmentation limit \citep{2012A&A...539A.148B}:

\begin{equation}
S t_{\mathrm{frag}}=0.12 \frac{v_{\mathrm{f}}^{2}}{\alpha c_{\mathrm{s}}^{2}}.
\end{equation}
From $S t(s_\mathrm{max})=S t_{\rm frag}$, we find $s_{\rm max} \propto \Sigma_{\rm g}^{-1}$. Particles smaller than 
this size can not grow to exceed it. They attain a coagulation-fragmentation equilibrium size distribution close to 
that of the MRN approximation. Particles larger than this threshold tend to coagulate to even larger sizes. 

The results of the fiducial model can be extrapolated to a more realistic model with a scaling factor $f_\Sigma$.
For example, the magnitude of $\Sigma_g$ for MMSN is everywhere ($f_\Sigma=$) 20-30 times higher than that in the 
fiducial model (Eqns \ref{eq:fiducial} \& \ref{eq:MMSN}).  For the same Stokes parameter, the corresponding value of 
$s_\mathrm{max}$ is scaled with the same factor (Eqn \ref{stokes} and the results of the modified model). If the gas and 
dust are supplied from the outer boundary with the same initial $\Sigma_d/\Sigma_g (\simeq 0.01,$ corresponding to the 
protostar's composition), the magnitude of $\Sigma_d$ would also be scaled by the same factor of $f_{\Sigma}$. The 
combined modification in both $s_\mathrm{max}$ and the total $\Sigma_d$ leads to the preservation of the $\sigma_d-s_d$ 
distribution as that of the fiducial model (Fig \ref{fig:DSD}) with a uniform increase in $\sigma_d (s_d)$ and a
uniform decrease in $\sigma_d (s_d) / \Sigma_g$ (by factor $< f_\Sigma$ for all sub-mm ($s_d \lesssim10^{-3}$m) grains. This would not affect the size distribution very strongly while
the opacity derived from our fiducial size-distribution is already consistent with previously used infrared wavelength averaged opacity tables for realistic environments (see \S 
\ref{sec:opacity} for our method to obtain opacity from the relative dust size distribution). Therefore, based on the modified model and the scaling relation, we extrapolate that the presence of an 
pebble-isolating super Earth (with $M_c=10 M_\oplus$) leads to a final enhanced opacity corresponds to our results of $\Sigma_d/\Sigma_g \vert_{\rm final}
\simeq 0.28$ (see Fig \ref{fig:opacity}, left panel).

In one additional simulation, a $M_c=20M_{\oplus}$ core is placed at 1AU inside a disk with same $\Sigma_g$ distribution
as the fiducial model.  The growth of total dust surface density $\Sigma_d$ around the ring/planet is shown in Fig 
\ref{fig:lineargrowth}.  This more massive planet induces a deeper partial gap in gas profile and less grains slip 
through.  The accumulation of dust and gas around the planet are significantly mitigated, albeit the growth of $\Sigma_d 
/\Sigma_g$ remains the same. In \S\ref{sec:4} we will discuss what this result implies for \textit{gas giant} 
formation.

\section{From Dust Size Distribution To Accretion Timescale}
\label{sec:3}
\subsection{Atmospheric Cooling Model}
In this section we apply the simulations results to determine the \textit{in situ} accretion timescale of a growing super Earth in passively irradiated regions at 0.1-1AU, 
and then indirectly extrapolate how similar effects may influence super Earths inside active regions. 
To model a planet's accreting process from solid core to runaway, we use the two-layer model developed by \citet{2014ApJ...786...21P}.

After the protoplanet achieves a steady core mass, the accreting process is primarily characterized by the change in the envelope's mass.  During 
the process of interest, the rocky core $M_c$ is surrounded by gas with increasing $M_{atm}$ until the process becomes 
unstable and the planet enters the runaway phase, which occurs when $M_{atm}\approx M_c$.  We determine the state (T, P, 
M) of the planet for a range of $M_{atm}=M_c\times \mathrm{GCR}$ in increasing order, and then deduce the $\Delta 
t$ that has elapsed between each snapshot by the calculated parameters. 

For every snapshot, we have a closed equation set:

\begin{equation}
\begin{aligned} \text{Mass Distribution: }&\frac{d M(<R)}{d R} =4 \pi R^{2} \rho_g \\ \text{Hydrostatic Equilibrium: } &\frac{d P}{d R}=-\frac{G M(<R)}{R^{2}} \rho_g \\ \text{Heat Transfer: }&\frac{d T}{d R} =\frac{T}{P} \frac{d P}{d R} \nabla \\ \text{Ideal Gas EOS: } &P=\dfrac{\rho_g}{\mu}\mathcal{R}T,
\end{aligned}
\label{structure}
\end{equation}
where $\mathcal{R}$ is the specific gas constant, while the temperature gradient is defined as
\begin{equation}
    \begin{aligned}
        \nabla=&\text{min}(\nabla_{ad}, \nabla_{rad}),\\
        \nabla_{ad}=& \dfrac{\gamma-1}{\gamma},\\        \nabla_{rad}=&\frac{3 \kappa P}{64 \pi G M \sigma T^{4}} L.\\
    \end{aligned}
\end{equation}
which changes discontinuously at the radiative-convective boundary (RCB). We assume that the luminosity of the planet is a global constant in each snapshot and is the eigenvalue of the equation set.

For mixed hydrogen and helium, it usually suffices to assume $\mu=2.35$ and $\gamma=1.4$ \citep[ACL20]{2014ApJ...786...21P}, however the result of \citet{2014ApJ...797...95L,2015ApJ...800...82P} 
considering the ionization of $\mathrm{H}$ shows that in reality $\gamma$ may be down to $\approx$1.25 depending on 
the dissociation fraction of hydrogen. It also depends on the degree of moisture in the planet atmosphere \citep{2019arXiv190209449A} and has a considerable impact 
on the evolution of super Earths. Taking into account these uncertainties, we consider 
both $\gamma=1.25$ and $\gamma=1.4$ for the \textit{innermost} convective zone in our simulation.

Given $T$, $P$ at the outer boundary, and $M$ at inner \& outer boundary $M_c$ and $M_c+M_{atm}$, we search by iteration 
the appropriate eigenvalue of luminosity by integrating the above planetary-structure equations from the outer boundary inward.  At the inner boundary, we impose the condition $M(R_{in})=M_c$.  Consistent with our fudicial numerical model, we adopt a 
core mass of $M_c=10M_{\oplus}$ with inner boundary $R_{in}=2.2R_{\oplus}$ in our cooling model. These values fall in the observed range from various exoplanet search campaigns.  For the outer boundary of
the envelope, $R_{out}$ is the smaller of the Hill and Bondi radius of the planet and it generally increases with $r$, which we set at typical values 0.1 and 1AU.  In the next section, we also consider the case of $20 M_\oplus$ core
located at 5, and 10 AU. 

\subsection{Evolution from Atmosphere to Envelope}

After solving the static structure functions for each GCR, we link these snapshots in time by calculating the time elapsed
between each adjacent snapshot with:

\begin{equation}
    \Delta t=\frac{-\Delta E+\left\langle e_{M}\right\rangle \Delta M-\langle P\rangle \Delta V_{\langle M\rangle}}{\langle L\rangle},
\label{eqn:time}
\end{equation}
Where $\langle X\rangle$ denotes the mean value of quantity $X$ in two adjacent snapshots, while $\Delta X$ is the difference between the snapshots. The terms in the numerator in Eqn \ref{eqn:time} account for a) the change in total energy:

\begin{equation}
    E=-\int [\frac{G M(<R)}{R}+u] d M, u=\mathcal{R}\left(\nabla_{\mathrm{ad}}^{-1}-1\right) T,
\end{equation}
integrated from the core to the RCB at radius $R_{\mathrm{RCB}}$; b) the energy obtained by accreting gas with specific energy

\begin{equation}
    e_{M}=-\left.\frac{G M(<R)}{R}\right|_{R_{\mathrm{RCB}}}+\left.u\right|_{R_{\mathrm{RCB}}};
\end{equation}
and c) the work done on the planet by the contracting envelope with $\Delta V_{\langle M\rangle}$ being the change in the volume enclosing the average of the convective masses of the two snapshots, and $\langle P \rangle$ evaluated at RCB. The surface terms turns out to be negligible except after runaway accretion, consistent with the findings of \citet{2014ApJ...797...95L,2014ApJ...786...21P}. For the initial snapshot, we define $t_{initial}=|E|/L$.




\subsection{Opacity Modification}
\label{sec:opacity}
In the previous section, we showed that small grains can effectively migrate through the pebble barrier after
pebble isolation has been achieved.  In order to
convert the perturbed dust size distribution into opacity, we generalize the fiducial scaling adopted by \citet{2014ApJ...789L..18O}. The total opacity is the sum of the gas and grain opacities:

\begin{equation}
    \kappa=\kappa_{\mathrm{gas}}+\kappa_{\mathrm{gr}}=\kappa_{\mathrm{gas}}+\sum_i \kappa_{\mathrm{geom}}(s_{i}) Q_{e}(s_i).
\end{equation}

The gas opacity which dominates at large temperature can be approximated by analytical expressions from \citet{1994ApJ...427..987B}. The grain geometric opacity $\kappa_{\mathrm{geom}}$ (per unit mixture, since it goes into the temperature gradient) and efficiency factor $Q_{e}$ for a species of grain with size $s_i$ are given by:
\begin{equation}
    \kappa_{\mathrm{geom}}(s_{i})=\dfrac3{4 \rho_{\bullet} s_i}Z_i, Q_e(s_i)=\min (0.3 x_i, 2),
\end{equation}
In our environment we choose $\rho_{\bullet}=1.5 \mathrm{g/cm}^2$ for silicate, ice and metal (generally referring to everything else) mixture with mass weight $1:6:1$, and $x_i=2 \pi s_i / \lambda_{\max }(T)$ where $\lambda_{\max}(T)$ is the peak wavelength of blackbody radiation from Wien's law. The specific value of $\rho_{\bullet} \sim 1$ does not affect the equilibrium dust size distribution very much since the Stokes number (Eqn \ref{stokes}) is globally modified within a factor of 3 in accordance with the internal density. In the disk environment the dust-to-gas ratio for a certain dust species is $Z_i=\sigma_d(s_i)/\Sigma_g$ where $\sigma_d(s_i)$ is its surface density yielding to $\sum \sigma_d(s_i)=\Sigma_d$. Therefore, our total grain opacity (cgs-units) is explicitly given by:

\begin{equation}
    \kappa_{\mathrm{gr}}=1.7\sum_{s_i<s_{crit}}\dfrac{ T\sigma_d(s_i)}{\Sigma_g}+0.5\sum_{s_i>s_{crit}}\dfrac{\sigma_d(s_i)}{s_i\Sigma_g},
\end{equation}
where $s_{crit}=\dfrac{1}{\pi T} \mathrm{cm}$ is the transition size.

\begin{figure*}[htp!]
\centering
\includegraphics[width=1\textwidth]{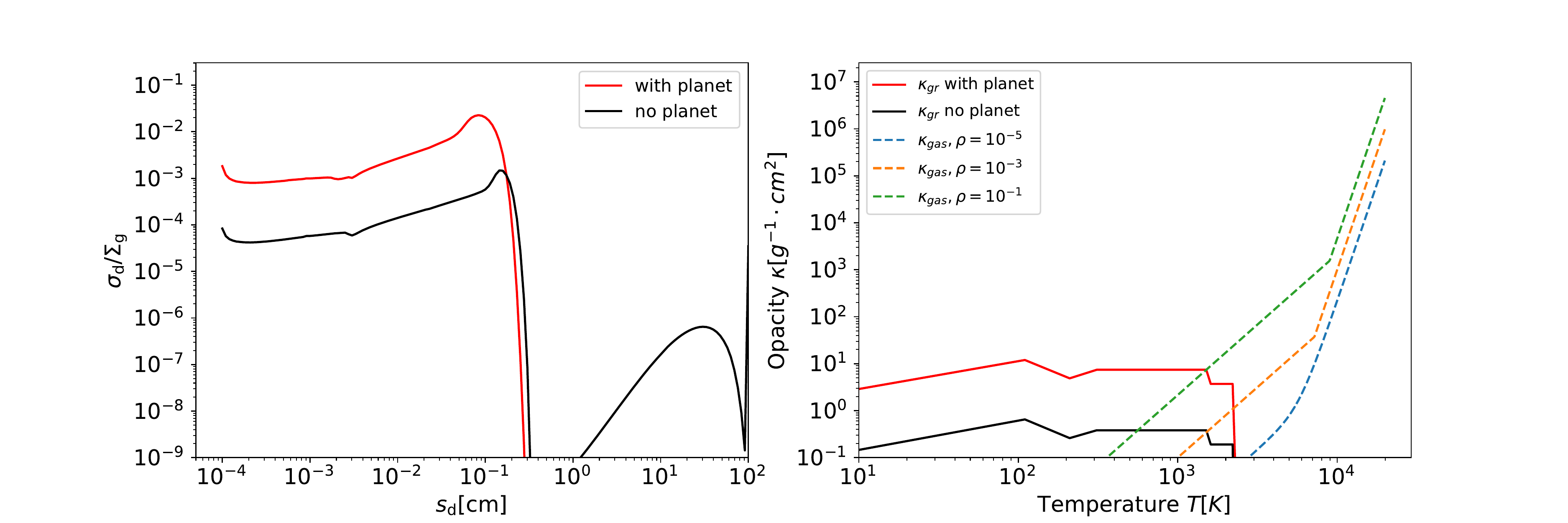}
\caption{Left panel: our adopted dust size distribution (normalized by gas surface density) at the planet location for unperturbed disk model (blue) and perturbed disk environment (black); Right panel: the corresponding relation of grain opacity $\kappa_{\mathrm{gr}}(T)$ for unperturbed and perturbed case, and also the gas opacity $\kappa_{\mathrm{gas}}$ for some characteristic values of $\rho_g$.}
\label{fig:opacity}
\end{figure*}

As we integrate the structure equations from the disk environment (at the boundary) to the RCB neglecting further coagulation, we can assume the densities $\rho_d(s_i)$ all scale linearly with gas density throughout the outer radiative zone, 

\begin{equation}
Z_i(R)=\left.\dfrac{\rho_d(s_i,R)}{\rho_g(R)}\right|_{R}\equiv\left.\dfrac{\sigma_d(s_i)}{\Sigma_g}\right|_{disk}
\end{equation}
so that our $\kappa_{gr}$ is only a function of temperature $T$. To approximate sublimation, we artificially take away the contribution of ice at 160K, silicate at 1600K, and essentially all the remaining metal at 2300K. In Fig \ref{fig:opacity} we plot out our adopted dust size distributions (normalized by gas density) for perturbed/unperturbed disk and the corresponding relation of $\kappa_{\mathrm{gr}}$ and $T$, and also $\kappa_{\mathrm{gas}}$ for some characteristic values of $\rho_g$. The opacity for perturbed disk is generally larger by a factor of 10-20 than that of the unperturbed disk.

It is worth mentioning that \citet{2014ApJ...789L..18O} adopted a representative-size approximate approach to study grain growth in a protoplanet's atmosphere. Assuming the dust size distribution at any $R$ is characterized by a single size $s$, they found that for monomer opacity sources (as opposed to porous agglomerates), a characteristic size of $10^{-4}$ cm at the boundary would grow to become $\sim 10^{-3}$ cm in most of the atmosphere and the dust-to-gas ratio would be reduced by a factor of $\sim 5$ compared to the boundary value. This suggestion led ACL20 to artificially reduce their grain opacity obtained from the table of \citet{1994ApJ...427..987B} by a factor to account for grain growth. 
For our unperturbed disk models (see below), our derived opacity from the self-consistent size distribution with a
fixed gas to dust ratio (Fig \ref{fig:opacity}) have already represented their values, as well as the ``dusty" opacity relation 
applied by \citet{2014ApJ...797...95L} extrapolated from \citet{2005ApJ...623..585F}, in which metals take the form of dust. Our analogous models for the unperturbed disk reproduce similar accretion timescales as previous cooling models (see below), while the effect 
of opacity reduction from dust sublimation and enhancement from collisional fragmentation were not considered in single-size approximation \citep{2014ApJ...789L..18O}. It is also unclear whether the representative-size approach could assert similar trends in the evolution dust-size spectrum of the disk region near the cores (Fig \ref{fig:DSD}).  Its
relatively flat $\sigma_d-s_d$ distribution with a broad peak around $s_d \sim 10^{-3}-10^{-2}$ cm provides a justification that grains
in this size range might provide the dominant opacity sources. 

Since we have incorporated the full coagulation effects in our computational method in \S\ref{sec:2}, the results of our 
simulation are found to be insensitive of the initial relative dust-size distribution across the disk as long as we maintain a steady 
injection from the disk's outer boundary. This verification supports the assumption that however an initial dust 
distribution might be centered around 1$\mu$m grains or cm-m-size pebbles, it would approach a similar size distribution, 
due to a coagulation-fragmentation equilibrium, when the disk gas and dust enter the outer boundary of the envelope, therefore we have a sound reason to neglect the dust-size evolution in the envelope.  This approximation is further supported if the heat transfer efficiency
through the envelope is primarily limited by the radiative zone near the envelope's outer boundary 
\citep{1996Icar..124...62P}.

While atmospheres of super-solar metallicity are consistent with IR observations of super Earths as GJ 1214b \citep{2013ApJ...775...33M}, GJ 436b \citep{2013ApJ...777...34M}, it raises the concern of affecting the mean molecular weight of the gas-dust mixture in the EOS (Eqns \ref{structure}). However, we note that as long as $\Sigma_d/\Sigma_g \lesssim$0.5, $\mu$ increases only weakly with the dust abundance \citep{2011ApJ...733....2N} so our enhanced dust distribution won't 
significantly affect the molecular weight. {Discovery of low-metallicity super Earths \citep{Bennekeetal2019} does not directly contradict our theory, since low-mass cores are allowed to undergo accretion in a relatively dust-free environment when the pebble reservoir is insufficient (See \S \ref{sec:5}, outstanding issue 2).}

In cases where we consider the contribution from entropy advection, we adopt the method of ACL20 by setting another 
boundary at 
$R_{\mathrm{adv}}=fR_{out}$. the region $R_{\mathrm{adv}}<r<R_{\mathrm{out}}$ is continually replenished with recycling 
disk gas on a faster timescale than the thermal timescale, with $f$ being typically $\sim 0.3$
\citep{LambrechtsLega2017}.  We force this region to be adiabatic by imposing 
$\nabla=\nabla_{ad}=2/7$, with no dependence on opacity. We adopt a conservative estimate of $f=0.4$.


\subsection{Passive Disk Models}
\label{sec:passivedisk}
After specifying boundary conditions and model parameters (including $M_c$, $M_\ast$ and $h$), two dominant physical 
effects: i) modification of opacity and ii) entropy advection, depend on a) energy source in the disk  
(viscous dissipation versus irradiation), b) channel of heat transfer in the disk (convection versus radiation), c) 
equation of state (value of $\gamma$) in the gaseous envelope, and d) distance from the central star.

We begin with a simplest prescription: a passively irradiated disk structure.  We adopt the boundary condition 
from the unperturbed values of an empirical MMEN model \citep{2013MNRAS.431.3444C,2014ApJ...797...95L} in which

\begin{equation}
\begin{aligned} \rho_{\mathrm{MMEN}}=6 \times 10^{-6}\left(\frac{r}{0.1 \mathrm{AU}}\right)^{-2.9} \mathrm{g} / \mathrm{cm}^{3} \\ T_{\mathrm{MMEN}}=1000\left(\frac{r}{0.1 \mathrm{AU}}\right)^{-3 / 7} \mathrm{K}. \end{aligned}
\label{eq:MMEN}
\end{equation}
An analogous Minimum Mass Solar Nebula (MMSN) is also used for the outer region of the disk in the discussions of \S\ref{sec:4}.  

\begin{figure*}[htp!]
\centering
\includegraphics[width=0.9\textwidth]{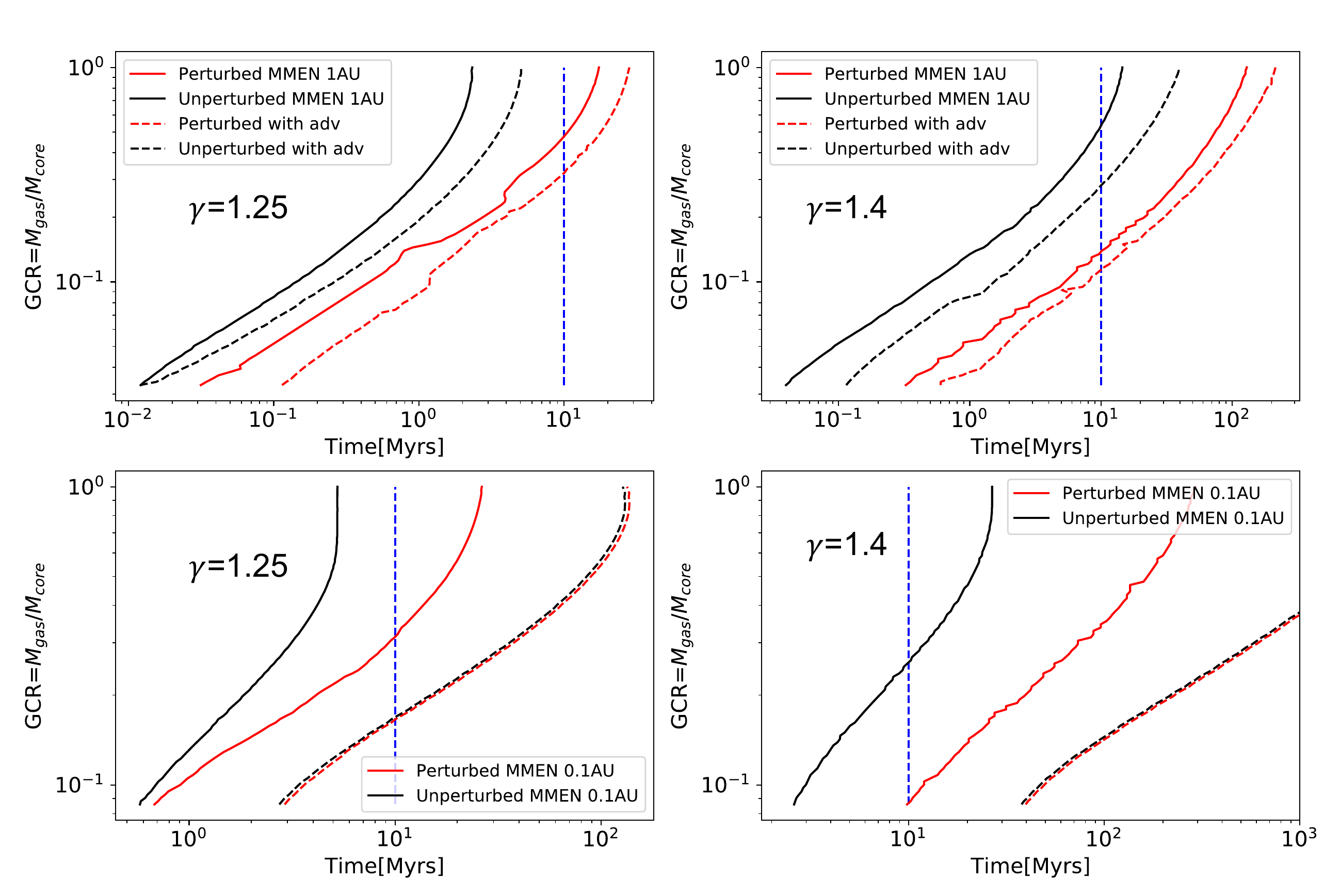}
\caption{The Gas-to-Core mass Ratio GCR as a function of time for a $10M_{\oplus}$ super-Earth core in passive disk regions. Upper panels are for 1AU in MMEN and lower panels for 0.1AU in MMEN. Left panels are for adiabatic index at inner convective boundary $\gamma=1.25$ and right panels are for $\gamma=1.4$. Black solid lines are for growth curves in unperturbed disk environments and red solid lines are growth curves considering the planet's perturbation on dust size distribution. The red and black dashed lines have taken into account entropy advection according to the method of ACL20. The blue dashed lines indicate the typical maximum disk lifetime of 10Myrs. At very early stages GCR $\sim$ 0.05, evolution is untraceable with the static snapshot method since the entire atmosphere is convective and independent of luminosity \citep{2014ApJ...797...95L}.}
\label{fig:passive}
\end{figure*}

The top and bottom rows of Fig \ref{fig:passive} show the evolution of a super Earth with core mass $M_c=10M_{\oplus}$ 
at $r=$1AU and $r=$0.1AU, respectively, in a passive MMEN environment.  We compare the outcome from the unperturbed (no 
planet, black lines) and the perturbed (with planet, red lines) disk models. The left and right panels represent, 
respectively, two values ($\gamma=1.25$ and $1.4$) for the adiabatic index of the \textit{inner} convection zone in the 
envelope.  For models which take into account entropy advection (red dashed lines with planet and blue dashed lines 
without planet), we assume $\gamma=1.4$ in the region $R_{adv}<R<R_{out}$. Models which neglect the entropy advection 
are represented with solid lines. 

For a super-Earth located at $\sim$1AU in a MMEN, the enhancement of opacity alone, (caused by pebble isolation) is 
able to prolong the runaway timescale by a factor of $\sim$10 and to prevent the transition from cores to gas giant planets.
A small increase in the value of $\gamma$ (from 1.25 to 1.4) have a similar magnitude effect. With $\gamma=1.4$, it is 
possible for the runaway timescale to exceed 10Myr without opacity enhancement associated with pebble isolation 
as shown previously by \citet{2013MNRAS.434..806B}. In general, the increase of accretion timescale by the opacity enhancement in the perturbed disk with $\gamma=1.25$ model is smaller than that with
$\gamma=1.4$ case. { This difference arises because the envelope is more susceptible to convection in the $\gamma=1.25$ case.}
Although the inclusion of entropy advection further doubles the runaway timescale, its contribution is overshadowed 
by the effect of opacity enhancement and increases in $\gamma$.

\begin{figure}[htp!]
\centering
\includegraphics[width=0.5\textwidth]{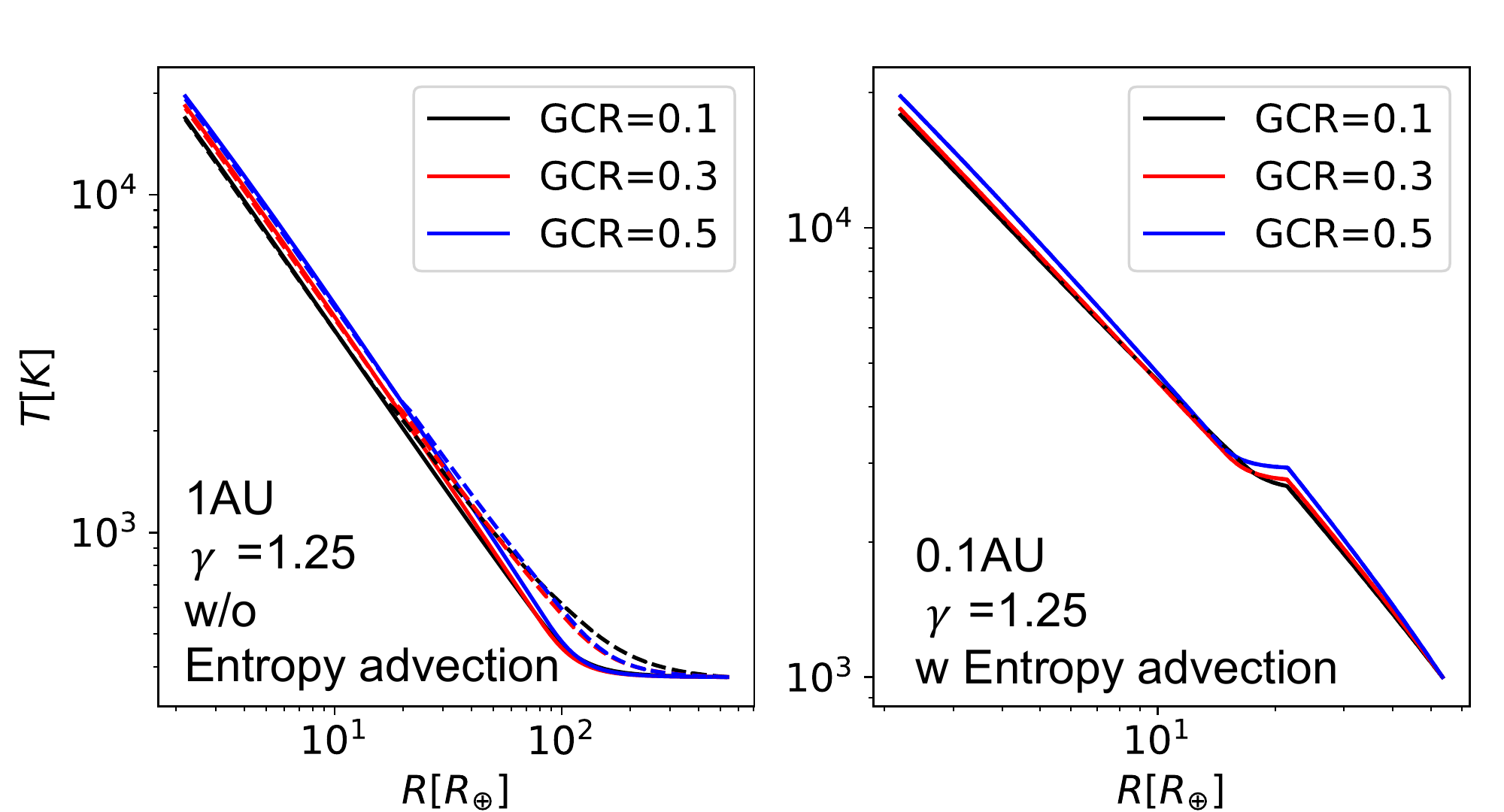}
\caption{{Atmospheric profiles for two exemplary scenarios in which grain opacity enhancement have very different impact. Left panel: $r=1$AU, $\gamma=1.25$, without entropy advection; Right panel: $r=1$AU, $\gamma=1.25$, with entropy advection. Bold lines are for unperturbed opacity and dashed lines are for perturbed opacity. The RCB marks the transition from radiative zone where the temperature gradient is typically shallow and the convective zone where gradient is steeper.}}
\label{fig:structure}
\end{figure}

For a super-Earth located at 0.1AU, the runaway timescale in unperturbed disks with $\gamma=1.25$ and $1.4$ 
is respectively less or greater than typical disk depletion time scale ($\lesssim 10$Myr).  For both values of 
$\gamma$, it is prolonged by the enhanced opacity when the effect of entropy advection is neglected. The ten-fold increase 
alone is adequate to prevent the transition to runaway gas accretion prior the severe depletion of the disk even for
$\gamma=1.25$.  This result is in agreement with previous studies \citep{2014ApJ...797...95L,2015ApJ...811...41L},
though with the perturbed disk model, we have provided a physical justification for the elevated metallicity 
during the post pebble-isolation stage.  Entropy advection further delays the 
onset of runaway accretion by more than another order of magnitude
as shown by ACL20. 
The fresh external supply of entropy also causes most of the $T \lesssim 2000$K outer zone to become convective (Fig \ref{fig:structure}, right panel).
In the radiative zone sandwiched between two convective layers, the temperature is mostly above the dust sublimation limit, therefore grain opacity enhancement has little influence on the atmospheric structure.

{In Fig \ref{fig:structure} we plotted the 1D temperature profiles in the growing envelopes, for one case where enhanced opacity has
a large impact ($r=1$AU, $\gamma=1.25$, without entropy advection) and another case for which enhanced opacity has little impact ($r=1$AU, $\gamma=1.25$, with entropy advection), for comparison. Solid lines are for unperturbed case, and dashed lines are for perturbed case. In the former case, the ten-fold increase in dust opacity is accompanied by a decrease in luminosity of similar magnitude to give rise to a mildly steeper temperature gradient in the outer radiative zone and a higher temperature close to the RCB. \footnote{For even larger metallicity, $\nabla_{rad}$ might increase so drastically such that the entire atmosphere become convective and runaway timescale is shortened.} In the latter case, the increase in metallicity does not affect the opacity/temperature at the RCB at all since it's already above sublimation threshold in the radiative zone.}

{The passive model used in this section is empirical and based on the assumption of stellar irradiation
as the main heating mechanism for the disk.  Despite the general agreements of various investigation, it is
prudent to examine their dependence on the model parameters.  Numerical simulations from 
\citet{2015A&A...575A..28B} showed that realistic PPD profiles undergo a gradual change from passive outer 
region to active inner region at around $\sim$1AU.   Within this transitional boundary, viscous heating dominates 
over stellar irradiation \citep{1980MNRAS.191...37L,2005A&A...442..703H,2007ApJ...654..606G} and
the disk thermal stratification and entropy profile are considerably altered. The results 
derived with the passive MMEN and MMSN disk model (see \S\ref{sec:4}) may be more applicable to regions outside rather than inside 1AU.}



\subsection{Active Disk Models}
\label{sec:activedisk}
ACL20 developed robust analytical models for radiative 
and convective PPDs and showed that at $0.1$AU, the growth of super Earths in an active disk differs significantly from 
that in a passive disk. The effect of entropy advection \citep{2015MNRAS.446.1026O,2015MNRAS.447.3512O} also becomes more
amplified. In the region heated by viscous dissipation, the high disk entropy quickly forces the envelope into a fully 
isothermal state at very early stages of core accretion and the envelope hardly grow at all.

ACL20's models are constructed with the opacity from unperturbed disk models.  In this subsection, we examine the 
contribution of elevated opacity in perturbed and active disks.
For simplification, we use the analytical models from \citet{1980MNRAS.191...37L}, ACL20.  In active disks where heat is generated through viscous dissipation and transferred by efficient 
convection, the thermal profile corresponds to an adiabat and 
\begin{equation}
    \begin{aligned} T_{disk, \mathrm{conv}}=& 391 \mathrm{K}\cdot r_{\mathrm{AU}}^{-27 / 28} \dot{M}_{-7.5}^{13 / 28} \alpha_{-2}^{-2 / 7} \\ & \times\left(\frac{M_{\star}}{M_{\odot}}\right)^{9 / 28}\left(\frac{\kappa}{\mathrm{cm}^{2} \mathrm{g}^{-1}}\right)^{2 / 7}, \end{aligned}
\label{eq:Tcon}
\end{equation}
where $r_{\mathrm{AU}}=r/1\mathrm{AU}, \dot{M}_{-7.5}=\dot{M} / 10^{-7.5} M_{\odot} \mathrm{yr}^{-1}$ and $\alpha_{-2}=\alpha / 0.01$. The density profile is 
\begin{equation}
    \begin{aligned} \rho_{g, \mathrm{conv}}=& 1.4 \times 10^{-11} \mathrm{g}/\mathrm{cm}^{3} \cdot r_{\mathrm{AU}}^{-87 / 56} \alpha_{-2}^{-4 / 7} \dot{M}_{-7.5}^{17 / 56} \\ & \times\left(\frac{M_{\star}}{M_{\odot}}\right)^{29 / 56}\left(\frac{\kappa}{\mathrm{cm}^{2} \mathrm{g}^{-1}}\right)^{-3 / 7}. \end{aligned}
    \label{eqn:conv}
\end{equation}

We assume $\gamma=1.4$ in the disk environment and artificially set the temperature to be below 2000 K to approximate the effect of dust sublimation, which significantly lowers the opacity and therefore the temperature gradient in the close-in region of the disk \citep{2001ApJ...553..321D}. This active disk model is significantly hotter and less dense with higher entropy than the passive MMEN model in the inner regions. For the active disks perturbed by an embedded core, we  \textit{assume} that the fragmentation of the trapped pebbles also increases the dust-to-gas ratio around the core by a factor of $\sim 10$. Since the grain opacity for an equilibrium size distribution around the planet is \textit{nearly} proportional to the dust-to-gas ratio (i.e. $\kappa_{\mathrm{gr}} \propto \rho_d/\rho_g$), the local opacity is also  enhanced by a factor of 10. We apply our perturbed/unperturbed opacity expressions from Fig \ref{fig:opacity} to check the difference in the gas accretion rate onto the core. From Eqn \ref{eqn:conv} we note that to keep our active disk model self-consistent with $\rho_g\propto (\rho_d/\rho_g)^{-3/7}$, $\rho_d$ has to increase by a factor of $\approx 3$ while $\rho_g$ decreases by a factor of $\approx 3$. The opacity also elevates the disk temperature accordingly. This effect increase the disk entropy $\propto \rho_g^{1-\gamma}T_{disk}$ by a factor of 3 for 1AU and 1.5 for 0.1AU (since the temperature is imposed to be below 2000K).

If heat transfer in the disk is primarily radiative, then
\begin{equation}
    \begin{aligned} T_{disk, \mathrm{rad}}=& 373 \mathrm{K} r_{\mathrm{AU}}^{-9 / 10} \alpha_{-2}^{-1 / 5} \dot{M}_{-7.5}^{2 / 5} \\ & \times\left(\frac{M_{\star}}{M_{\odot}}\right)^{3 / 10}\left(\frac{\kappa}{\mathrm{cm}^{2} \mathrm{g}^{-1}}\right)^{1 / 5} \end{aligned}
\label{eq:Trad}
\end{equation}
and
\begin{equation}
    \begin{aligned} \rho_{g, \mathrm{rad}}=& 1.7 \times 10^{-10} \mathrm{g} \mathrm{cm}^{-3} r_{\mathrm{AU}}^{-33 / 20} \alpha_{-2}^{-7 / 10} \dot{M}_{-7.5}^{2 / 5} \\ & \times\left(\frac{M_{\star}}{M_{\odot}}\right)^{11 / 20}\left(\frac{\kappa}{\mathrm{cm}^{2} \mathrm{g}^{-1}}\right)^{-3 / 10} \end{aligned}.
\end{equation}
In this region, in order for the dust-to-gas ratio to increase by a factor of 10, $\rho_d$ has to increase by a factor 
of $\approx 5$ while $\rho_g$ decreases by a factor of $\approx 2$. These changes would increase the disk entropy $\propto \rho_g^{1-\gamma}T_{disk}$ by a factor of 2 for 1AU and 1.3 for 0.1AU.

\begin{figure*}[htp!]
\centering
\includegraphics[width=0.9\textwidth]{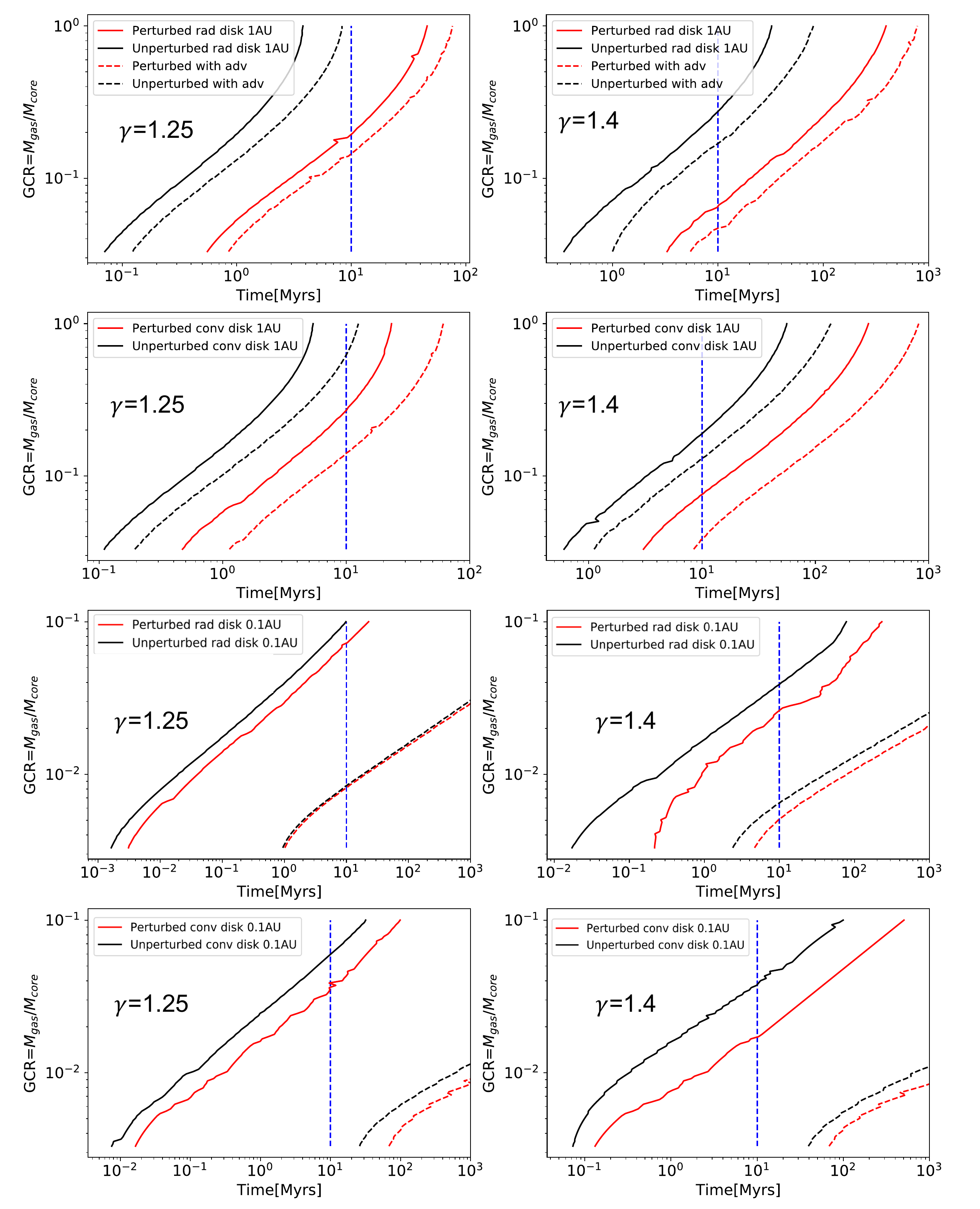}
\caption{The Gas-to-Core mass Ratio GCR as a function of time for super-Earth in active disk regions. Top two panels for a $10M_{\oplus}$ core at 1AU and lower two are for a $10M_{\oplus}$ core at 0.1AU (with \textit{0.1} as the upper limit of GCR). Left panels are for adiabatic index at inner convective boundary $\gamma=1.25$ and right panels are for $\gamma=1.4$. Black lines are for growth curves in unperturbed disk environments and red lines are growth curves considering the planet's perturbation on dust size distribution. The dashed lines are growth curves taken entropy advection into consideration.}
\label{fig:active}
\end{figure*}

Fig \ref{fig:active} shows the evolution of a super Earth at 1AU and 0.1AU in both radiative and convective environment. We compare the evolution of GCR for both the unperturbed and perturbed case, with and without entropy advection. Left panel is for $\gamma=1.25$ and 
right panel is for $\gamma=1.4$ (the adiabatic index for \textit{inner} convective zone). At 1AU, the temperature and gas density environment of active disks is not very different from the MMEN.  The enhanced opacity effect (and the change of boundary entropy as well) plays a more significant role than entropy advection in prolonging the runaway timescale. 

At 0.1AU, a much higher temperature and lower gas density render the timescales 2 orders of magnitude longer than the results from the MMEN. These findings are consistent with those of ACL20.  Contribution from the enhanced grain opacity becomes less influential since the local temperature is already approaching the sublimation limit of grains, and would be further quenched by entropy advection.
Interior to this region, gas becomes the opacity source. Nevertheless, we show that a 10$M_{\oplus}$ core 
embedded in an active disk at 0.1 and 1AU cannot make a transition to runaway gas accretion within 10 Myr. This 
result is robust for different values of $\gamma$, with or without entropy advection, as long as we take into account the opacity enhancement. Same can be said for the results of passive disks.

\subsection{Choice of Core Mass}
\label{sec:isomass}
Based on the observed properties of known close-in super Earths, we adopt $M_c=10M_{\oplus}$ 
for cores at 0.1 and 1AU in all the models presented above.  Similar core masses ($5-10M_{\oplus}$) 
have been obtained from classical theory \citep{1982P&SS...30..755S} and adopted in previous 
core-accretion studies \citep[ACL20]{2014ApJ...797...95L,2017ApJ...850..198Y}.  In the pebble
accretion scenario, the cores' growth is stalled and grain opacity is enhanced in the post pebble 
isolation stage.  It is more realistic to match $M_c$ with the pebble isolation mass. Numerical 
simulations on the onset of pebble isolation \citet{2014A&A...572A..35L} 
yield a critical core mass

\begin{equation}
M_{\rm Iso} \approx 20\left(h/0.05\right)^{3} m_\ast {M}_{\oplus}
\label{19}
\end{equation}
where $m_\ast=M_\ast/M_\odot$. The luminosity of T Tauri stars usually scales as $L_\ast / L_\odot \simeq m_\ast^2$ \citep{DAntonaMazzetilli1994}.  Under the assumption 
$T_{\rm MMEN}$ 
in Equation (\ref{eq:MMEN}) is proportional to  $L_\ast^{1/4}$, we find $
    h_{\rm MMEN} \simeq 0.02 r_{\rm 0.1 AU} ^{2/7} m_\ast^{-3/20} $ and that

\begin{equation}
    M_{\rm Iso, MMEN} \simeq 1.3 r_{\rm 0.1AU}^{6/7} m_\ast^{11/20} M_\oplus
\label{eq:passiveiso}
\end{equation}
for the MMEN models of the passive irradiated disks.  For the actively, viscously 
heated disks, the observed accretion rate \citep{1998ApJ...495..385H, Nattaetal2006, Manaraetal2012, Darioetal2014} ${\dot 
M}_\ast \propto m_\ast$ during the T Tauri phase.  
From Equations (\ref{eq:Tcon}) and (\ref{eq:Trad}), we find $h_{\rm conv} \propto
r^{1/56} m_\ast^{-3/28}$ and $h_{\rm rad} \propto r^{1/20} m_\ast^{-3/20}$ so that

\begin{equation}
    M_{\rm Iso, c} \propto r^{3/56} m_\ast ^{19/28} \ \ \ {\rm \&} \ \ \ M_{\rm Iso, r} 
    \propto r^{3/20} m_\ast ^{11/20}
\label{eq:activeiso}
\end{equation}
for the fully convective and radiative regions of the disk respectively.  Depending 
on the magnitude of disk's ${\dot M}$, $\alpha$ and $\kappa$, the normalization value
for these isolation mass is $\sim 10-20 M_\oplus$ over a wide disk region (0.1-1 AU).  
Since ${\dot M}$ declines during the depletion of the disk, both $M_{\rm Iso,c}$ 
and $M_{\rm Iso, r}$ also decreases with time.

Equation (\ref{eq:passiveiso}) indicates that the aspect ratio $h$ increases with $r$ for a core embedded in a passive 
disk heated by stellar irradiation. The limiting $M_{\rm Iso}$ also has a considerable dependence on $r$ and cannot 
assume to be uniform. For the MMEN model with a solar-type star, the 
isolation mass is $9M_{\oplus}$ at 1AU but $1.3M_{\oplus}$ at 0.1AU. However, to match with the observations
with the passive model, most 
super-Earths at 0.1AU must have acquired a rocky core with $M_c \simeq 5-10M_{\oplus}$ somehow, despite the termination 
of supplement from pebble accretion in the post pebble isolation stage \citep{2019A&A...630A..51B}. Cores with 
$M_c (\sim 10M_{\oplus})$ larger than $M_{\rm Iso, MMEN}$ may emerge from planetesimal coagulation within 0.1AU 
in a sufficiently solid-rich MMEN, provided that dynamical friction 
by other planetesimals or type I damping by residual gas does not quench such coagulation beyond an early 
embryo-isolation phase \citep{2013MNRAS.431.3444C}.  They may also gain additional mass from 
the mergers of pebble-isolating cores at the last stages of disk evolution \citep{2016ApJ...817...90L} 
or have migrated inwards from a larger distance \citep{LiuOrmelLin2017}. Regardless whether it can 
acquire $M_{\rm Iso}$ at the beginning of gas accretion phase, the runaway timescale would be further
\textit{prolonged} for a core with $M_c < 10 M_\oplus$, and render the avoidance of runaway robust.

We have already indicated in the previous section that passive models may not be appropriate for the inner 
regions of the disk where active viscous dissipation rather than passive stellar irradiation is the primary heating
mechanism.  In active disks, cores can acquire $M_c \sim M_{\rm Iso, c}$ or $M_{\rm Iso, r}$ through the 
accretion of inwardly migrating pebbles even though a relatively limited local reservoir of planetesimals 
may be inadequate for them to coagulate directly into super Earth embryos.  The supply of pebbles is 
interrupted after their isolation. Equation (\ref{eq:activeiso}) indicates that $h$ and $M_{\rm Iso}$ 
are very weakly increasing functions of $r$ in both the radiative and convective regions of the disk.  
Since the core's gas accretion is impeded by the opacity enhancement after they have enter the pebble 
isolation stage, the asymptotic mass of their atmosphere can only be minor fraction of $M_c$.  
This inference is consistent with the estimates on the super Earths' limited atmospheric loss due to post-formation 
photo-evaporation \citep{OwenWu2017}.  Moreover, core attain similar asymptotic mass $M_c \sim M_{\rm Iso, r}$
or $M_{\rm Iso,c}$ (both $\sim 10 M_{\oplus}$) over a wide range of distance (0.1-1AU). This extrapolation is
also in agreement with the observed mass and size similarity between multiple super Earths around common host 
stars \citep{Weissetal2018,Wu2019}.

\section{Transition to gas giants in the outer disk regions}
\label{sec:4}

In our attempt to account for the \textit{avoidance} of super Earth's runaway, we also find that the opacity enhancement 
effect seems to be in danger of quenching the emergence of gas giants from rocky cores altogether throughout the 
disk. In this section we present some additional results of accretion timescales for progenitor cores at outer regions of their natal disks (at $r \gtrsim 1$ AU), and discuss some mechanisms that recover their transition to gas giants. Since these are the regions where most gas giants are found around mature stars, we assume they have not migrated extensively from the sites where their progenitor cores made their runaway transition.

\subsection{Passive Disk Regions}
\label{passive4.1}
{At $r\gtrsim5$ AU, disk becomes more transparent with relatively low $\Sigma_g$.  As the stellar
irradiation become the dominant heating mechanism, the passive disk model may be more appropriately 
applied. }In this case, $h$ increases with $r$ as required in the models of the spectral 
energy distribution of PPDs \citep{ChiangGoldreich1997}. 

\begin{figure}[htp!]
\centering
\includegraphics[width=0.5\textwidth]{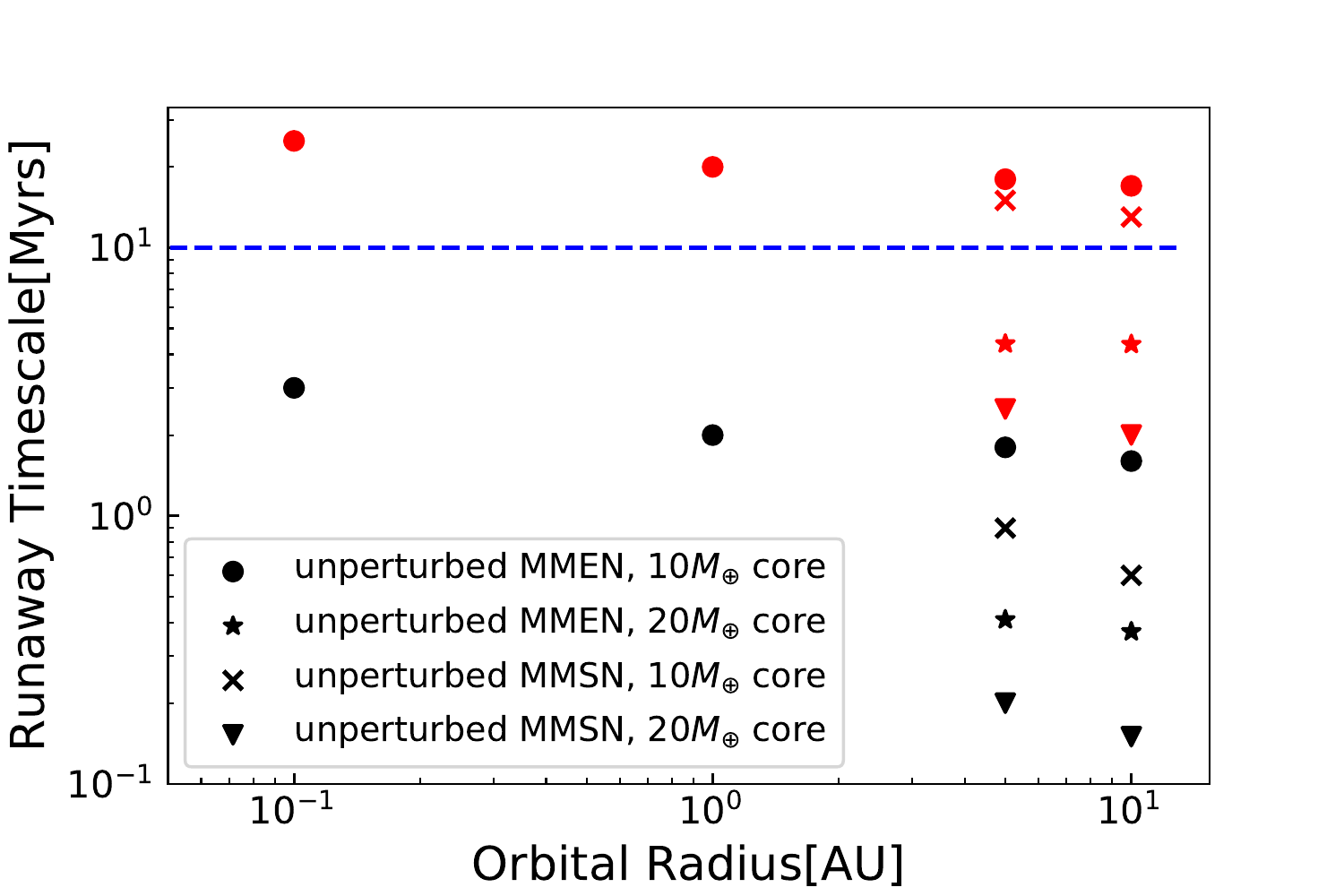}
\caption{A summary of runaway timescales in passive disks, for different core mass and in different opacity environments for $\gamma=1.25$, \textit{without} considering entropy advection. The runaway timescales for the unperturbed MMEN is insensitive to the orbital radius as consistent with \citep{2014ApJ...797...95L}. The blue dashed line indicates the disk lifetime. Black signs indicate runaway timescales for unperturbed diskenvironments  and  the  corresponding  red  signs  are  for  perturbed cases. We show that the most plausible way for a pebble-isolating gas giant embryo at 5-10 AU to \textit{recover} runaway is to have a core mass larger than $10M_{\oplus}$.}
\label{fig:times}
\end{figure}

\citet{2014ApJ...797...95L} found that the runaway timescale of $M_c = 10M_{\oplus}$ cores in a passive MMEN is 
insensitive of the orbital radius throughout the disk {for ``dusty" atmospheres. Similar results are found by \citet{HoriIkoma2011}.}  As an equivalence test, we reproduce their results with an 
analogous model for a $M_c=10M_\oplus$ core embedded in unperturbed passive MMEN disks for the $\gamma=1.25$ case without entropy advection (Fig \ref{fig:times} black circles). We confirm a runaway timescale $\sim 2$ Myr, which is insensitive to the radial distance, albeit the applicability of the passive model without entropy 
advection for $r \sim 0.1$AU remains questionable. 

For $r \gtrsim 1$AU, the runaway timescale is mostly determined by the magnitude of opacity in the radiative outer 
regions of the envelope and that entropy advection does not prolong it by more than a factor 2 (see 
\S\ref{sec:passivedisk}).  But, in regions perturbed by cores with pebble isolation mass, the grain 
opacity is elevated in the envelopes and the runaway time scale is lengthened beyond the disk lifetime of 
10Myrs (Fig \ref{fig:times} red circles), quenching gas giant formations via core accretion.

The MMEN model is an empirical prescription motivated by the observed distribution of close-in super Earths.
This model may not be well suited for the outer regions because $T_\mathrm{MMEN}$ reaches the water-ice sublimation temperature at 10 AU, well outside the location where asteroids are predominantly composed of volatile ices.  
In the construction of solar-system formation theory, the outer solar nebula is often 
approximated with a Minimum Mass Solar Nebula (MMSN) \citep{1977Ap&SS..51..153W,1981PThPS..70...35H}. As a 
comparison with the MMEN, we use an updated version of the MMSN model \citep{2010AREPS..38..493C} adopted by 
\citep{2014ApJ...786...21P} with profile

\begin{equation}
    \begin{aligned} \Sigma_{\mathrm{MMSN}} &=70 (\dfrac{r}{10AU})^{-3 / 2} \mathrm{g\cdot} \mathrm{cm}^{-2} \\ T_{\mathrm{MMSN}} &=45 (\dfrac{r}{10AU})^{-3 / 7} \mathrm{K}, \end{aligned}
\label{eq:MMSN}
\end{equation}
that's 3 times colder than the MMEN everywhere and offers more cooling.  If we use the standard opacity for the unperturbed passive MMSN 
model, runaway accretion for a $M_c=10 M_\oplus$ core at 5 and 10 AU would occur within a few Myr quicker than that obtained with the unperturbed MMEN model. In this case, it is easier to reach runaway accretion before the severe depletion of the disk gas (Fig \ref{fig:times} black crosses). 
However, if the production of sub-mm size grains is elevated by the collisions of accumulating pebbles in their barriers, the cooling in the envelope would still be subdued by the enhanced opacity effect. The colder gaseous 
environment in the MMSN also expands the radiative zone in the envelope and therefore amplifies the influence of 
the higher opacity. If the opacity is enhanced (Fig \ref{fig:times} red crosses), the runaway timescale for a $M_c=10 M_\oplus$ core 
at 5AU in a cold MMSN would still be longer than the disk depletion timescale.
 
\subsection{Pebble Isolation Mass in Outer Disk Regions}
In all the above models, we have adopted $M_c=10 M_\oplus$ to match that of known super Earths.  This value of
$M_c$ is $\sim M_{\rm Iso, r}$ or $M_{\rm Iso, c}$ at $r= 0.1-1$AU in an active disk (see \S\ref{sec:isomass}). 
Based on the findings 
that the runaway timescale has much stronger dependence on core mass than its surrounding boundary conditions 
\citep{2014ApJ...797...95L,2015ApJ...811...41L}, \citet{2016ApJ...817...90L} proposed that super Earths can effectively 
\textit{avoid} runaway if the core mass stays small within most of the disk lifetime. Conversely, cores with relatively 
large $M_c$ can \textit{reach} runaway transition and evolve into gas giants. 

Since cores accrete pebbles until they impose pebble isolation, they would attain $M_c \sim M_{\rm Iso}$ if there is an 
adequate supply of inwardly migrating pebbles.  Cores with $M_c \gtrsim 10 M_\oplus$ can emerge in the {passive outer regions of the disk where $M_{\rm Iso}$ is an increasing function of distance}. For a passive disk with the MMEN model, Eqn \ref{eq:passiveiso} indicate that $M_{\rm Iso, MMEN}$ can reach $\sim 70 M_\oplus$ at $r=10$AU around solar type stars.  For the less extreme passive MMSN models,
\begin{equation}
    h_{\rm MMSN} \simeq 0.05 r_{\rm 10 AU} ^{2/7} m_\ast^{-3/20},
\end{equation}
\begin{equation}
    M_{\rm Iso, MMSN} \simeq 20 M_\oplus r_{\rm 10AU}^{6/7} m_\ast^{11/20}.
\label{eq:isommsn}
\end{equation}
At sufficiently large distance (5 and 10 AUs), especially around more massive solar-type stars ($M_\ast \sim M_\odot)$, $M_{\rm Iso, MMSN}$ may exceed the typical core mass $M_c=10M_{\oplus}$ we have so far adopted. 

\subsection{Transition From High-mass Cores to Gas Giants}
\label{sec:passgasgiant}
The potential prospect of  cores with a relatively large $M_c$ raises the possibility of reaching runaway accretion 
prior to the severe depletion of the disk. For example, a $M_c=20M_{\oplus}$ core in a passive MMSN has a 
runaway timescale of $\sim$ 2Myrs even after considering an opacity enhancement similar to the $10M_{\oplus}$ case (red wedges in Fig \ref{fig:times}). The runaway timescale for this 
core in a passive MMEN is only a factor of $\sim 3$ longer (stars). These timescales are less than the disk lifetime. 

There are several potential pathways for the core to acquire a large $M_c$ initially.

\noindent
1) In the relatively thin 
($h \lesssim 0.05$) irradiating disk regions, although $M_{\rm Iso, MMSN} (\lesssim 20 M_\oplus)$ sets a limit 
on $M_c$ that is insufficient for
individual cores to enable the transition to runaway accretion, isolated pebbles accumulate at their migration 
barrier until their local $\Sigma_d$ is sufficiently large to promote the formation of later-generation cores 
with $M_c \sim M_{\rm Iso, MMSN}$.  Under this circumstance, the requirement of a larger core mass for rapid transition 
to runaway gas accretion may be attained through the mergers of oligarchic protoplanetary cores 
\citep{1998Icar..131..171K,2004ApJ...604..388I,2013ApJ...775...42I} rather than solely via the monolithic 
orderly accretion of migrating pebbles. Since the stochastic merger events require dynamical instability among 
the oligarchic embryos and cores, the emergence of gas giants is a probabilistic process which generally leads 
to diverse kinematic properties. 

\noindent
2) In the MMSN model, $h \gtrsim 0.05$ at $r \gtrsim 10$AU. In such regions where $M_{\rm Iso, MMSN} \gtrsim
20M_{\oplus}$, cores formed out of planetesimal coagulation with $M_c < M_{\rm Iso}$ 
continue to accrete pebbles until either they have acquired a relatively large pebble-isolation mass (and enhanced opacity) 
or the disk runs out of building block pebbles before severe gas depletion. In the latter case, the grain opacity retains its value 
for the unperturbed disk such that cores with $M_{c} \gtrsim10 M_\oplus$ may also be able to reach runaway (see Fig \ref{fig:times}), if there is still sufficient gas to accrete. This is more likely to happen in the MMEN where an extreme pebble isolation mass of $M_{\rm Iso, MMEN} \gtrsim35M_{\oplus}$ is obtained at $r \gtrsim 5$AU.

\noindent
3) The magnitude of $M_{\rm Iso, MMSN}$ also increases with the stellar mass $M_\ast$ 
(Eqn \ref{eq:isommsn}). Solar mass stars generally have cores with larger pebble isolation mass than K and M
dwarfs. The probability for embryos to merge is also higher around more massive host stars \citep{Liuetal2016}. Both 
effects suggest that the formation of gas giants may be more prolific in disks surrounding \textit{high-mass} stars.
This extrapolation is consistent with the observed correlation between the occurrence rate of gas giants with the
mass of the host stars \citep[e.g.][]{2013A&A...549A.109B}. 

\noindent
4) For the $\gamma=1.4$ case \footnote{{We emphasize again that this is a slowest extreme case neglecting hydrogen dissociation or  vapor condensation in  the  envelope in the inner convective zone \citep[][ACL20]{2014ApJ...786...21P}. The typical central temperature of 5-10 $M_{\oplus}$ cores is $\sim 10^4$ K which exceeds the dissociation limit of 2500K, the $\gamma=1.4$ scenario is only for reference.}}, the 
runaway timescale for any given $M_c$ are prolonged further (albeit this dependence on $\gamma$ is 
less sensitive compared to that at smaller radii because the convective zones make up less fraction 
of the atmosphere). A core mass of $M_c \sim 35M_{\oplus}$ is needed to enable the transition to runaway 
accretion in perturbed passive disks. With the required $M_c > M_{\rm Iso, MMSN}$, merger of multiple
pebble-isolated cores is necessary for the transition to runaway gas accretion in the MMSN. 


\subsection{Active Disk Regions}
In young and globally accreting disks, it's still possible that transition region between active and passive disks extend between 1-10AU \citep{2007ApJ...654..606G}, allowing for potential gas giant formation in an active environment.

\begin{figure}[htp!]
\centering
\includegraphics[width=0.5\textwidth]{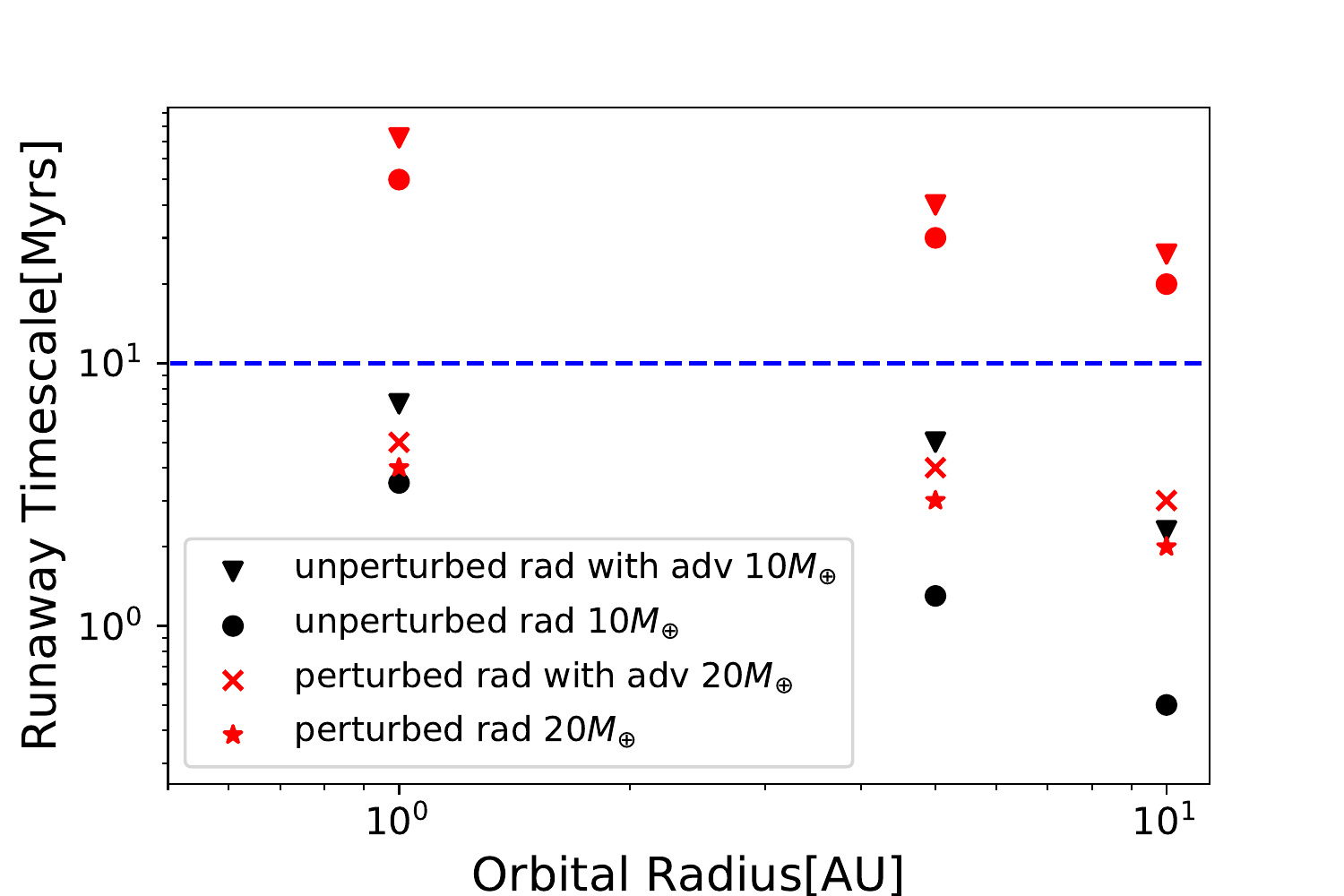}
\caption{A summary of runaway timescales in radiative disks, for different core mass and in different opacity environments for $\gamma=1.25$, with and without considering entropy advection. The blue dashed line indicates the disk lifetime. Black signs indicate runaway timescales for unperturbed disk environments and the corresponding red signs are for perturbed case.}
\label{fig:timeactive}
\end{figure}

The accretion of $M_c=10M_{\oplus}$ rocky cores at 0.1AU is much slower in active disks due to higher temperature and 
lower density profile than the MMEN (See \S\ref{sec:3}).  This tendency is reversed in the outer regions at $r\gtrsim 
5$AU where the active disk models are denser and colder and the accretion is faster than passive disk models. In Fig 
\ref{fig:timeactive} we plot a summary of runaway timescales at $r \gtrsim 1$AU regions within the \textit{radiative} 
disk (those in convective disks are similar), with $\gamma=1.25$ in inner convective zone. For the \textit{unperturbed} 
disk regions, the retardation contribution associated with entropy advection is recovered at larger radii (black 
circles and wedges).  Our result is
consistent with the findings of ACL20. However, the enhanced opacity for the perturbed disks is still able to prolong 
the runaway timescale to become much longer than 10Myrs while quenching the influence from entropy advection (red 
circles and wedges), in contrast to the models for 0.1AU where entropy advection dominates over opacity enhancement.

Nevertheless, transition to runaway accretion can still be reached for a $M_c = 20M_{\oplus}$ core (Fig \ref{fig:timeactive} red crosses and stars), while we need $M_c \sim 35M_{\oplus}$ for $\gamma=1.4$ models. For these active disk models, $h \lesssim 0.05$ at 5-10AU with insufficient pebble isolation mass ($M_{\rm Iso, c}$ and $M_{\rm Iso, r} < 20 M_\oplus$).  
The discussions in the previous section are still applicable, i.e. the progenitor cores of gas giant planets may acquire their critical masses for transition to runaway gas accretion either from probabilistic mergers of oligarchic embryos, or through the monolithic accretion of pebbles at relatively large distances (5-10 AU) around their solar-type or more massive host stars.

\subsection{Coexistence of Super Earths, Gas and Ice Giants}
\label{4.5}
Protostellar disks are primarily heated by viscous dissipation within $\sim 1$AU and
stellar irradiation outside a few AU such that active disk models are appropriate for
$r \lesssim 1$AU and passive MMSN models are more applicable to $r \gtrsim 5$AU.  
The discussions in \S\ref{sec:3} show that super Earths are robustly retained in the inner active 
disk regions, especially in the proximity of their host stars.  Consideration in 
\S\ref{sec:passgasgiant} also indicate that gas giant cores can form through the stochastic
merger of pebble-isolating cores in the outer passive disk regions. By extrapolation,
we anticipate the formation of long-period gas giants to be associated with the emergence
of close-in super Earths.  This inference is consistent with the observed occurrence
correlation \citep{2018AJ....156...92Z,Bryanetal2019} with the Solar System being an exception.

At larger distance ($\sim 5$AU) where $M_{\rm Iso, MMSN} \gtrsim 10 M_\oplus$, the runaway timescale for 
pebble-isolated cores in perturbed disks is comparable to or shorter than the disk depletion timescale. 
As its envelope accretes a significant fraction of its initial $M_c$, a growing planet clears 
out an increasingly deep gap in $\Sigma_g$ around its orbit.  Less amount of sub-mm-size grains is able 
to slip through the steepening pressure gradient outside the gap. This effect is evident in the 
simulation we have presented in \S\ref{sec:2} with a $M_c=20M_\oplus$ core. However, a deepening gap in gas profile still leads to greater metallicity and the grain opacity (per unit mass mixture) growing towards $\sim 10$ times the unperturbed value (Fig \ref{fig:lineargrowth}, dashed lines). This effect would continue to keep the gas accretion at a pace lower than normal (which proves calculation of prolonged timescales in \S\ref{sec:passgasgiant} necessary), albeit the core is still able to reach runaway. As the planet's mass grow exponentially post-runaway, its tidal torque on the disk intensifies. Eventually, gas accretion is quenched 
by the formation of a much deeper gap in $\Sigma_g$ when the planet reaches Jupiter-mass \citep[e.g.][]{1986ApJ...309..846L,Dobbs-Dixonetal2007}. For fixed grain opacity scaling/metallicity, a reduction of $\sim$ 1000 in gas density is required to significantly quench accretion rate further.

At larger radii ($\gtrsim 10-20$AU), disks may run out of building block pebbles for cores. Yet, the 
$M_{\rm Iso}$ continues to increase with $r$ to values $>20 M_\oplus$. If mos disk gas is nearly depleted by the time the pebbles run out, cores which fail to accrete 
sufficient $M_c$ to undergo transition either to pebble isolation or runaway gas accretion are retained 
as ice giants \citep{2014A&A...572A..35L,2018A&A...612A..30B}.  Without pebble isolation, they are not 
embedded among an enhanced population of sub-mm grains with elevated opacity. These ice giants may be able 
to accrete a non-negligible gaseous envelope, similar to Uranus and Neptune in the Solar System. 

\section{Summary and Discussion}
\label{sec:5}

In this paper, we attempt to address the probability of transition from super Earths to gas giants through runaway gas 
accretion.  The technical issue we examine is whether a $\sim 10M_{\oplus}$ progenitor core's pebble isolation effect can influence its own core 
accretion process. We carry out simulations using simple PPD models with dust coagulation starting from an overall 
dust-to-gas ratio of 0.01.  We find that as the dust-to-gas ratio in the accumulating pebble barrier builds up to $\sim 
1$, the stalled and accumulated pebbles undergo collisions to produce much smaller grains which bypass the migration 
barrier.  The dust-to-gas ratio around the vicinity of a $10M_{\oplus}$ core reaches $\sim 0.2$, more than ten times the
value at the same location in the steady state of a PPD without the planet.  The dust distribution around the planet is made up of small particles 
since larger pebbles are blocked at the pressure maxima. This process significantly increases the grain opacity in the outer 
radiative zone of the envelope surrounding the core (see Fig \ref{fig:opacity}, left panel), which self-limits 
the planet's growth rate.

Since the dynamics of pebble isolation is generic, we suggest that similar evolutions of normalized dust distribution at 
the planet location also take place in a variety of denser and more realistic PPD models, and at different locations. Based on this 
assumption, we compared the \textit{in situ} gas accretion process of super Earths with a core mass of 
10$M_{\oplus}$ located at $\lesssim1$AU. This value for core mass is chosen reconciling the observation results and the pebble isolation mass at $\lesssim1$AU. We adopt both unperturbed disk models in different heat transfer conditions and those with 
the realistic opacity environments perturbed by the planet.  We adopt a generalized formula of grain opacity which is 
appropriate with the dust's size distribution in the atmosphere. 

For the short-period super Earths, we consider many models at 0.1 AU for unperturbed and perturbed disks.
In addition to the grain opacity, these models are based on two different values of $\gamma$. For the 
outer boundary condition, we also adopt the passive (heated by stellar irradiation) phenological MMEN models 
as well as the active (heated by viscous dissipation) fully convective and radiative models for the disk.  
In all but one model, the timescale for a $M_c=10 M_\oplus$ core to reach runaway gas accretion is longer 
than 10 Myr.  The only model which led to runaway gas accretion is obtained with the 
opacity of an \textit{unperturbed} disk, a low $\gamma$ equation of state, and the temperature distribution from 
the passive MMEN passive model.  Since the active disk models as well as the perturbed opacity environment are more realistic for the inner disk
region, short-period super Earths are robustly preserved with a low-mass initial atmosphere in the 
post pebble isolation phase.  By itself, the enhanced opacity, due to sub-mm grains produced by the 
trapped pebbles, can greatly suppress the efficiency of heat transfer and limit the gas accretion rate.  
In an active disk, the pebble isolation mass is $\sim$ a few to 10 $M_\oplus$ and it is a very weakly 
increasing function of the disk radius. Multiple super Earths may emerge from the trapped pebbles 
with similar masses, radii, and separation around common host stars.  These extrapolations are consistent with 
the observed mass and period distribution of close-in super Earths.

We also considered several cases at 1 AU. If we adopt the opacity for the unperturbed disk and $\gamma=1.25$ at the inner convective zone, the planet would undergo 
transition to runaway accretion within $\sim 2-3$ Myr, in both active and passive disk environment.  However, if we take into account the opacity enhancement after 
pebble isolation, the runaway accretion timescales would be increased by a factor of 10 in accordance with analytical 
results from \citet{2000ApJ...537.1013I}.  This change would be enough to prevent runaway for the cases we have 
considered and naturally account for the high occurrence rate of super Earths found by microlensing surveys \citep{2012RAA....12..947M}.

In previous attempts to address the issue of super Earths retention, the potential contribution from entropy advection 
has been suggested \citep{2015MNRAS.446.1026O,2015MNRAS.447.3512O}.  In this paper, we revisit the issue of entropy advection with the 
method constructed by ACL20.   Our results indicate that at 1AU, although this effect does delay
the onset of runaway gas accretion by a factor $\sim 2$, it makes much less difference than the 
enhanced opacity effect.  At $r=0.1$AU, the continuous replenishment of disk gas penetrates into the core's Hill sphere 
and mixes entropy with gas within a significant fraction of the envelope. Since this recycling zone is adiabatic, it effectively suppresses dependence in the grain opacity by pushing the radiative zone into interior regions. In contrast to $r=1$ AU, entropy advection in the proximity of the host star reduces the cooling efficiency and further slow down the gas accretion rate by more than an order of magnitude.
Nevertheless, since the enhanced opacity alone already prevents the cores from transforming into gas giants, we reach  
the robust conclusion that residual super Earth cores are effectively retained at $r \lesssim 1$AU. 

For $10M_{\oplus}$ rocky cores even further out in passive disk regions beyond $r\gtrsim 5$AU 
the effect of entropy advection is negligible while the enhanced opacity in perturbed disks is still able 
to prolong the accretion timescale from $\sim 2$Myrs to $\sim 20$Myrs and quench the emergence of gas giants. 
Nevertheless, runaway accretion can be readily reached for $M_c \gtrsim 20M_{\oplus}$ progenitor core. 
Such cores may form via pebble accretion in relatively hot passive disks provided there is an adequate pebble 
reservoir, or through  probabilistic mergers of two or more less massive cores. Both processes are more 
probable around at the outer regions of disks around high mass stars. This inference is consistent with 
the observed dependence of gas giant occurrence rates on their host stars' mass and common coexistence 
of long-period gas giants and close-in super Earths. 

There are several outstanding issues. 1) Our analysis is focused on grain population consisted of monomers 
in both the hydrodynamic simulations, and the atmospheric cooling model.
In the discussion of opacity expressions, we reconcil our opacity relations with the coagulation results of 
monomers from \citep{2014ApJ...789L..18O}, and the opacity relations adopted by previous works 
\citep[ACL20]{2014ApJ...797...95L}. However, \citet{2014ApJ...789L..18O} also showed that coagulation would reduce the 
opacity more drastically if grains mostly exist in the form of porous agglomerates, which would bring about a ``low 
$\kappa$ catastrophe" that quickly triggers runaway. How would the interaction between gas and \textit{porous} dust 
during pebble isolation, which follows a different dynamical regime, influence the accretion of progenitor cores appeals
to further research. {{On the other hand, we did not take into account evaporated dust above the sublimation temperature that raises the gas-phase metallicity in the atmosphere, which might delay the envelope growth further up to $Z \sim 0.5$, beyond which the increased mean molecular weight would expedite the collapse of the envelope \citep{HoriIkoma2011}.}}

2) {We adopt the conjecture that the runaway coagulation growth and pebble accretion of planet embryos are quite efficient.} 
In the outer regions of disks
around massive stars, it is possible for embryos to emerge a pebble isolation mass $\gtrsim 10 M_\oplus$ 
\citep{2013ApJ...775...42I}.  In this case, pebble accretion is not needed for the cores to reach runaway before 
the disk is severely depleted.  In disks with modest ${\dot M}$ around sub-solar-mass stars, embryos emerge with 
embryo-isolation mass $\lesssim$ a few $M_\oplus$.  We adopt the assumption that they can quickly evolve into 
cores with pebble isolation mass through the continual accretion of fast streaming pebbles \citep{2010A&A...520A..43O,2018A&A...612A..30B} {within a few Myrs, so they still have time to reach runaway accretion post-pebble isolation}.
In our analysis on the necessary condition for super Earth retention versus gas giant formation, we also
assume the depletion time scale of pebbles is comparable to or or longer than that of gas \citep{2014ApJ...780...53C,LambrechtBistch,2018A&A...612A..30B}. Under
these assumptions, the opacity enhancement effect is sustained over the course of gas accretion on to the cores. If the 
pebbles are depleted when the gas is tenuous,  planets' growth would still be stalled and the failed cores would be preserved 
as {low-metallicity} super Earths in the inner regions or evolve into ice giants in the outer region. {If the the pebble accretion is too inefficient for gas giant embryos in the outer regions, they are more likely to run out of pebbles than reach pebble isolation mass before $\sim $ Myr \citep{linetal2018}. Its gas accretion would be at the normal rate unperturbed by enhanced opacity and it could still quickly evolve into a gas giant. More detailed time dependent analysis of growing cores in evolving disks with a limited supply of pebbles can be considered in future studies.}

3) We neglect the possibility of Types I and II migration.  {\it In situ} formation remains a feasible possibility 
that can reproduce super Earths' observed distributions of orbital periods, mass, as well as mutual inclinations
\citep{2013ApJ...775...53H}. But, migration timescales may be relatively short, especially near the 
host stars where many super Earths are found. The direction and destiny of migration remain unclear due to the 
uncertainties in the saturation of co-rotation torques/horseshoe drags under different disk conditions.  There are 
various mechanisms that could halt the migration process \citep{2012ARA&A..50..211K,2013LNP...861..201B}.For their asymptotic properties, migration may not matter very much if the cores' migration is stalled at some traps.  {Furthermore, lack of a pile-up in the orbital period distribution of super Earths are observation evidence that large-scale migration may not have occurred \citep{LeeChiang2017}.}
Over a wide active region of the disk (0.1 to a few AU), $M_{\rm Iso, c}$ and $M_{\rm Iso, r}$
is $\sim 10 M_\oplus$ (see \S \ref{sec:activedisk}). If their $M_c$ growth is quenched by pebble isolation (i.e. their $M_c 
\sim M_{\rm Iso, c}$ or $M_{\rm Iso, r}$, \S \ref{sec:isomass}), migration throughout this region would not significantly change 
the high retention probability of super Earth since their runaway time scale would be much longer than the disk 
depletion time scale.  In the outer regions of the disk ($\gtrsim 5$AU), the runaway timescale for the cores 
and the probability of gas giant formation with similar $M_c$ do not depend significantly on their location (Fig. 
\ref{fig:timeactive}).  Therefore, adding migration might not qualitatively change our conclusions. 
Nevertheless, migration should be taken into account in the determination of the asymptotic kinematic 
distribution of super Earths and gas giants.

{4) Our hydrodynamical simulation for a passive disk is carried out for gas viscosity $\alpha_g=10^{-3}$. The calculations of specific isolation mass (\S \ref{sec:isomass}, \S\ref{passive4.1}) are also done in this limit applying the formula from \citet{2014A&A...572A..35L} for viscosity of the same order. A generalized expression from \citet{2018A&A...612A..30B} $M_{{\rm Iso}}=5 M_{\oplus}\left(\dfrac{h}{0.03}\right)^{3}\left[0.34\left(\dfrac{-3}{\log _{10} \alpha_g}\right)^{4}+0.66\right]$ adds another viscosity-related factor to the original scaling. For larger/smaller viscosity, the power law dependence of isolation mass with respect to $r$ is still similar to \S \ref{sec:isomass}, only the specific values would be larger/smaller by a global factor, similar to the influence of a larger/smaller host star mass. In nearly inviscid disks $\alpha_g \sim 10^{-4}$, the scaling law could deviate further from Eqn \ref{19} to have a steeper dependence on orbital radius which also helps to account for the dichotomy between outer gas giants and inner super Earths \citep{FungLee2018}, the pressure bump induced by a super Earth is smaller but more asymmetric structures would appear in the dust profile \citep[][see their Fig 1]{2017ApJ...843..127D}. How does the opacity enhancement effect perform in such disk regions are left for future work.}

5) We also neglect the self-gravity of the gas/dust in the simulations.  This approximation is justified around location
of the planet where the disk is not strongly perturbed. However, our simulations do show that inside the pebble barrier,
$\Sigma_d/\Sigma_g$ can grow to order unity, which might trigger streaming instability \citep{2005ApJ...620..459Y} and 
lead to pebble aggregation.  Under this condition, the self-gravity of the large aggregates may play a large part in 
generating new planetesimals and cores. Alternatively, cores can form directly out of gravitational instability 
\citep{2004ApJ...608.1050G, 2014ApJ...780...53C}. This process may recur and lead to the sequential formation of 
multiple close-in super-Earths. Such a scenario has been suggested for the TRAPPIST-1 system \citep{2017Natur.542..456G}
by \citet{2018ApJ...868...48K,2019ApJ...879L..19K}.  Since the accumulation is post-natal, multiple super Earths may
emerge in disks with dust density much smaller than the MMEN.  In \S \ref{sec:passgasgiant}, we suggest the possibility that
dynamical instability may lead to the merger of two or more super Earths with sufficient mass to undergo transition 
to runaway accretion and the formation of gas giant at $r \gtrsim 5$AU. Applying the full coagulation codes to study this separate mechanism as well as the opacity environment of the emerging new planets, would be the subject of another paper.

\acknowledgments
This project was motivated by a discussion with Yanqin Wu.
We thank her, Eve Lee, Mohamad Ali-Dib, Andrew Cumming, Eugene Chiang, Chris Ormel, Shengtai Li, Shigeru Ida,
and Songhu Wang for helpful discussions. Y.X.C. thanks the Department of Astronomy of UC Berkeley for its hospitality since part of this project was completed during his visit in UC Berkeley, supported by Tsinghua Xuetang Cultivation Program. He also thanks Y.P.L. and H.L. gratefully acknowledge the support by LANL/LDRD. 
This research used resources provided by the Los Alamos National Laboratory Institutional Computing Program, which is supported by the U.S. Department of Energy National Nuclear Security Administration under Contract No. 89233218CNA000001.

\textit{Softwares}: LA-COMPASS \citep{2005ApJ...624.1003L,2009ApJ...690L..52L}, Numpy \citep{2011CSE....13b..22V}, Scipy \citep{2019arXiv190710121V}, Matplotlib \citep{2007CSE.....9...90H}.


%






\bibliography{sample63}{}
\bibliographystyle{aasjournal}



\end{document}